\documentstyle[prd,aps,eqsecnum,preprint]{revtex}

\begin{document}

\title{Flow Equations for Gluodynamics in the Coulomb Gauge}

\author{Elena Gubankova,
Chueng-Ryong Ji, and Stephen R. Cotanch}

\address{Department of Physics, North Carolina State University,
Raleigh, NC 27695-8202}


\maketitle

\begin{abstract}
A systematic procedure to consistently formulate a field theoretical,
QCD bound state problem with a fixed number of constituents is outlined.
The approach entails applying the Hamiltonian flow equations, which are
a set of continuous unitary transformations,
to a QCD motivated Hamiltonian with a confining interaction. The
method is developed in detail for gluodynamics in the Coulomb gauge
to obtain an effective block-diagonal Hamiltonian appropriate to a reduced
Fock space with fixed number of dynamical gluons.
Standard many-body techniques are used to numerically diagonalize this
Hamiltonian in a constituent two gluon Fock space.  The calculated gluon
condensates and glueball masses are in good agreement with QCD sum
rule and lattice results.
\\\\
\hspace{-1.5em}PAC number(s): 12.39.Mk, 12.39.Ki, 11.10.Ef, 11.10.Gh
\end{abstract}


\section{Introduction}

Quantum Chromodynamics (QCD) is  generally regarded as the theory governing
hadronic structure and interactions. This broad acceptance follows, in
part, from the success of high energy perturbative calculations where
asymptotic freedom prevails. However, a significant
gap still exists in implementing QCD in the
non-perturbative, low energy regime where physical observables have
only been described by lattice calculations, QCD sum rules and
phenomenological models. Our study addresses
this shortcomming and details a new approach which directly incorporates
canonical QCD at high energies and phenomenology, including confinement,
>from low energy, to produce a realistic effective Hamiltonian appropriate
for bound state investigations.

Because the relevant hadronic degrees of freedom do not clearly
follow from the canonical QCD Hamiltonian, we look to phenomenological
models, which are reasonably developed, for initial guidance.  Of
particular interest is gluonic structure, the focus of this work,
which has been described by vastly alternative approaches:
strings \cite{nielson}, flux tubes \cite{patonisgur}, stochastic gluon
configurations \cite{simonov} and constituents
\cite{barnes} confined by a bag \cite{jaffe} or potential
\cite{eichten,ssjc96}.  In this paper we continue to build from our
previous work \cite{ssjc96,flesrc,robertson} within
the constituent picture, again viewing both quarks and gluons as dressed
partons subject to an effective interaction.  We assume this interaction is
dominated by a confining potential whose explicit form is not essential,
provided it is long ranged and generates bound states. Our effective
Hamiltonian also contains perturbative interactions, including radiative
corrections, governing the ultraviolet (UV) region.  These terms are directly
and unambiguously obtained from the leading order coupling constant
expansion of the canonical Hamiltonian.

With these preliminaries we now address the essence of this paper, a
systematic method for consistently solving a generic, field
theoretical QCD motivated Hamiltonian.  The method we present is
based upon the flow equations suggested by Wegner \cite{wegner}
and applied to QED within the light-cone framework
\cite{gubweg,gubpaulweg}.  Because a field theoretical
solution involves diagonalization in an infinite
Fock space with arbitrary number of quanta, an exact solution is not
possible.  Instead we seek a finite Fock space formulation and begin by
dividing the
complete Fock space into two components, a tractable $P$ subspace spanned by
states with a small number of quanta and the remainder $Q =1 - P$. This
leads to a Hamiltonian matrix to be diagonalized of the form

\begin{eqnarray}
   H = \left(
   \begin{array}{cc}
      PHP & PHQ \\
      QHP & QHQ
   \end{array}\right)
\,.\label{eq:i3}\end{eqnarray}
Next we use the flow equation method to block diagonalize this matrix
yielding the effective Hamiltonian

\begin{eqnarray}
   H_{ eff} =\left(
   \begin{array}{cc}
     PH_{ eff}P & 0            \\
     0             & QH_{ eff}Q
   \end{array}\right)\,.
\end{eqnarray}
One can then separately diagonalize the two blocks which are now
uncoupled.  It is important to note that the flow equation approach is
significantly different than the Tamm Dancoff approximation (TDA)
even though they both may use the same $P$ space.
In the TDA the off-diagonal blocks $PHQ$ and $QHP$ are eliminated
simply by truncation, whereas in the flow equation method
they are eliminated by transformation to $PH_{eff}P$ and $QH_{eff}Q$.

In the flow equation approach the Hamiltonian is unitary
transformed continuously
\begin{equation}
   H(l)=U(l)H(0)U^{\dagger}(l)
\,,\end{equation}
\begin{equation}
   U(l)=T_l {\rm exp}\, \left(\int_0^l \eta(l')dl'\right)
\,,\end{equation}
where $l$ is the continuous (flow) parameter ($l=0$ corresponds to the
initial Hamiltonian, Eq. (\ref {eq:i3})), $T_l$ the
$l$ ordering operator and $\eta$ the generator of the transformation.  Wegner's
flow equations are then an expression of this unitary transformation in
differential form
\begin{equation}
   \frac{dH(l)}{dl} = [\eta(l), H(l)]
\,,
\end{equation}
with the specific generator choice
\begin{equation}
\eta(l) = [H_d(l), H(l)]
\,,
\end{equation}
where $H_d$ is the diagonal part of $H$ and $H_{eff} = H_d (l \rightarrow
\infty)$.  Using this generator, it is shown in section II how the off-diagonal
terms $QHP$, which connect Fock states with different particle number, can be
eliminated as $l \rightarrow \infty$ and included in the block diagonal
effective
interaction.  Hence coupling  to the
$Q$ space is incorporated into the $P$ space and represented by new
terms in the effective Hamiltonian.
Once $PH_{eff}P$ is determined, the bound state problem can be investigated in
detail by direct diagonalization.  In general, $PH_{eff}P$ will become more
complicated as the $P$ space becomes smaller which may necessitate additional
approximations in obtaining the effective Hamitlonian.  Hence the strategy
it to find a balance between size of the $P$ space, which is determined by
computational effort and resources, and level of approximations.  In the
current work we will apply this approach to gluonia and restrict the $P$
space to
Fock states with at most two dynamical (dressed) gluons.

In the next section we review the
flow equation approach for a general Hamiltonian.  It is important to
note, however, that the solution of these equations is
quite involved. This is because the equations are nonlinear and coupled and
also because the choice of generator needed to decouple the $P$ and $Q$
spaces generates complicated many-body interactions.  For most
realistic applications, therefore, the flow equations will be implemented
iteratively. This has been
demonstrated in a variety of investigations such as condense matter problems
\cite{And}, \cite{spBos},
electron-phonon \cite{LeWe} and positronium \cite{gubweg} treatments,
the latter two involving perturbative expansions in coupling constant.  Here we
adopt a slightly different  perturbative scheme and
invoke the weak coupling assumption commonly used in nuclear structure
many-body calculations.  There one introduces a mean field, or
Hartree-Fock component of the interaction, and expresses the Hamiltonian
interaction as a sum of this term and a residual force, the difference
between the original potential energy and the mean field.  An approximate
effective interaction is then generated by treating the residual
interaction perturbatively.  The physical nuclear many-body
states are finally obtained by diagonalizing this effective interaction in the
$P$ space spanned by eigenfunctions of the mean field Hamiltonian.
The success of the method is determined by choice of
model space $P$ and phenomenological interactions.  Although our problem
is not a central field situation, we nevertheless can utilize this
concept since a dominant mean field is present in the form of the
phenomenological confining interaction.  This directly governs the
quasi-particle (dressed) gluon basis functions which
in turn form the many-body basis spanning the $P$ space.  We
identify our perturbative interaction as the difference
between the original QCD interaction and the phenomenological
confining potential and approximate this residual interaction at high
energies by the leading order canonical Hamiltonian interaction.
Again the overall utility of this method will depend significantly upon
choice of $P$ space and phenomenological input.
For example in our pure gluon application we assume that
three and higher gluon Fock states only couple weakly
to the $P$ space.  Because the dressed gluon mass is of order 1
$GeV$, which sets the scale for a large energy denominator in a
perturbative expansion, this approximation should be reasonable.
However, it may develop in quark sector applications that the weak
coupling approximation is not sufficient for low Fock space components
such as exotic 4 quark ($qq\bar q \bar q$) and hybrid ($q\bar q g$)
states.  In such cases the
$P$ space must be expanded to explicitly include these higher components
and this
is  a subject for future investigations.

Finally we mention the issue of renormalization which can also be accomodated
by the flow equations.  Because the complete Fock
state expansion contains states of arbitrarily large energies,
UV divergences occur for certain matrix elements.  However, since QCD is
a renormalizable theory, these UV diverging terms can be
isolated, regularized and then by introducing counterterms,
renormalization can be achieved perturbatively order by order in
the coupling constant
\cite{robertson,gw93}.  As demonstrated in the following sections, the flow
parameter, actually $l^{-1/2}$, plays an additional role as an UV cut-off
in the flow equation elimination of the off-diagonal terms.
Hence, the flow equations embody the renormalization concept of Wilson
wherein the high energy modes are integrated out in the path
integral representation yielding a low energy effective action.
Our method is also analogous to the Glazek and Wilson similarity
transformation scheme \cite{gw93} with a continuous cut-off function
\cite{robertson}.  In particular, as shown in section IV.A, we obtain
a well behaved gap equation with canonical counterterms cancelling
the leading UV divergence.  This is one reason why we do not incorporate the
phenomenological confining terms in the implementation of
the flow equations since
they are not renormalizable.  Further and related, including the
confining potential in the flow generator would produce additional
effective interaction terms that are still phenomenolgical.
Since an initial phenomenological interaction is already present, the model
framework remains unchanged and we therefore bypass this step.

This paper is divided into five sections and three appendices.  In the
next section we summarize the general flow equation method,
highlighting the key issues. In section III we
apply the flow
equations and specify the explicit form of the
model QCD Hamiltonian.  In addition to the weak coupling motivation discussed
above, we also wish to disentangle all complexities from chiral symmetry
breaking and thus focus on pure gluodynamics.
This section also developes and presents our final effective Hamiltonian.
Numerical results are then given in section IV, with the gap equation,
gluon constituent mass and condensate analyzed in section IV.A and
glueball bound states and masses in section IV.B.  Finally,
results and future work are discussed in the concluding section
along with an extensive appendix containing full details of
our comprehensive treatment.

\section{The Flow Equation Method}

We now summarize the important elements of
flow equations for a general Hamiltonian. We follow
Wegner \cite{wegner} who first formulated this approach
to construct an effective  bound state Hamiltonian
appropriate for field theories.
Since this method does not depend on the
specific system or detailed nature of the Hamiltonian, it is
beneficial to first conceptually outline the general idea
before our application in section III
to a complicated field theory.
See ref.
\cite{gubpaulweg} for complete details.

Continuing with Eqs. (1.3-1.6), the first step is to divide the
Hamiltonian, $H(l)$, into diagonal and "rest", or off-diagonal,
components, $H_{d}(l)$ and $H_{r}(l)$, respectively
\begin{eqnarray}
   H_d(l) =\left(
   \begin{array}{cc}
     PH(l)P & 0     \\
     0      & QH(l)Q
  \end{array}  \right)
\,,\end{eqnarray}
\begin{eqnarray}
  H_{r}(l)=H(l)-H_d(l)=\left(
  \begin{array}{cc}
    0      & PH(l)Q \\
    QH(l)P & 0
  \end{array} \right)
\,.\end{eqnarray}
It is precisely the off-diagonal part, which connects Fock states with
different particle number, that we wish to remove.  Next, and
most important, is the choice of generator.  As shown by Wegner our
block diagonalization goal can be acheived by choosing
\begin{eqnarray}
   \eta(l) = [H_d(l), H(l)]
\,,\label{eq:i5}\end{eqnarray}
which is off-diagonal by direct evaluation
\begin{eqnarray}
   \eta(l)=\left(
   \begin{array}{cc}
     0          & P\eta(l)Q  \\
     Q\eta(l)P  & 0
   \end{array} \right)
\,,\end{eqnarray}
with $P\eta Q = PHPHQ - PHQHQ$ and similar result for $Q\eta P$ upon
interchange of $P$ and $Q$.
The flow equation, Eq. (1.5), can then be written for the important
sector Hamiltonians $PHP$ and $PHQ$
\begin{eqnarray}
   \frac{d}{dl} PHP &=& P\eta QHP - PHQ\eta P
\,,\label{eq:i8}\\
   \frac{d}{dl} PHQ &=& P\eta QHQ - PHP\eta Q
\,.\label{eq:i9}\end{eqnarray}
The essence of the flow equation approach is that with Wegner's generator
choice the absolute value of the off-diagonal terms, $PH(l)Q$, $QH(l)P$,
monotonically decreases to zero as the flow parameter tends to infinity.
This can
be directly shown by taking the trace of their product and computing the change
with $l$ yielding,  using EQ. (2.6), a negative derivative
\begin{eqnarray}
    \frac{d}{dl} {\rm Tr} PHQHP
    &=& {\rm Tr} \Big( P\eta Q (QHQHP - QHPHP)\Big)
\nonumber\\
    &+& {\rm Tr} \Big((PHPHQ - PHQHQ) Q\eta P \Big)
\nonumber\\
    &=& 2 {\rm Tr} (P\eta Q\eta P)
\nonumber\\
    &=& -2 {\rm Tr} (P\eta Q(P\eta Q)^\dagger)\leq 0
\,,\label{eq:i11}\end{eqnarray}
where the last step follows from the antihermiticity of $P\eta Q$.
Since the trace of $PHQHP$ is positive definite,
$P\eta(l)Q\rightarrow 0$ in the limit $l\rightarrow\infty$.
Hence, $H_d$ commutes with $H$ and the effective Hamiltonian, $H_{ eff}$,
approaches complete block diagonal form with increasing flow parameter.

We can formally solve the flow equations to provide further detailed
insight and also the starting point for our gluodynamics application
in the next section.   The Hamiltonian, $H = H^{free} + H^I$, consists of
free (non-interacting), $H^{free}$, and interacting, $H^I$, components.
We then work in a fixed basis spanned by eigenfunctions of the free sector
Hamiltonians $PH^{free}P$ and $QH^{free}Q$ with eigenvalues $E_p$ and $E_q$,
respectively.  Here the indices $p$ and $q$ denote all states (both Fock and
energy) in the respective $P$ and $Q$ spaces. Representing the matrix elements
of transformed Hamiltonians $PH(l)P$, $PH(l)Q$, etc. in this basis by
$h_{pp'}(l)$,
$h_{pq}(l)$, ... the flow equations take the form
\begin{eqnarray}
   \frac{d h_{pp'}}{dl} &=&
   \sum_q \Big( \eta_{pq}h_{qp'}-h_{pq}\eta_{qp'} \Big)
\,,\\
\frac{dh_{pq}}{dl} &=& \sum_{q'}  \eta_{pq'}h_{q'q}- \sum_{p'}
h_{pp'}\eta_{p'q}
\,,\\
\eta_{pq} &=& \sum_{p'}  h_{pp'}h_{p'q}- \sum_{q'} h_{pq'}h_{q'q}
\,,\end{eqnarray}
where all matrix elements are functions of $l$.
The initial conditions (flow parameter $l = 0$) for the matrix elements
of the sector Hamiltonian $PH(0)P$ are,
$h_{pp'}(0)=E_p(0)\delta_{pp'}+\delta
E_p^{I}(0)\delta_{pp'}+h_{pp'}^{I}(1-\delta_{pp'})
\equiv E_p^{I}(0)\delta_{pp'}+h_{pp'}^{I}|_{p\neq p'}$,
and for the Hamiltonian $PH(0)Q$,
$h_{pq}(0)$.  Here $\delta E_p^{I}$ and $h_{pp'}^{I}|_{p\neq p'}$ are the
diagonal
and off-diagonal matrix elements of the interaction Hamiltonian $H^I$ in
the $P$
space.  We
now separate the diagonal and off-diagonal matrix elements in
Eqs. (2.8-2.10)
\begin{eqnarray}
\frac{dE_p^{I}}{dl} &=& \sum_q \Big( \eta_{pq}h_{qp}-h_{pq}\eta_{qp} \Big)
\,,\\
   \frac{d h^{I}_{pp'}}{dl} &=&
   \sum_q \Big( \eta_{pq}h_{qp'}-h_{pq}\eta_{qp'} \Big)
\,,\\
\frac{dh_{pq}}{dl}   &=& -\Big( E^I_p-E^I_q \Big) \eta_{pq}
  + \sum_{q' \neq q}  \eta_{pq'}h^I_{q'q}-
\sum_{p' \neq p} h^I_{pp'}\eta_{p'q}
\,,\\
   \eta_{pq}   &=& \phantom{-}\Big( E^I_p-E^I_q \Big) h_{pq} +
\sum_{p' \neq p}  h^I_{pp'}h_{p'q}- \sum_{q' \ne q} h_{pq'}h^I_{q'q}
\,.\end{eqnarray}
This system of coupled equations can be approximately decoupled
by assuming that the
off-diagonal $PHQ$ sector matrix elements
are small as compared to diagonal $PHP$ elements, i.e.
$|h_{pq}(0)|\ll |h_{pp'}(0)|$ (from the flow equations this is also valid
for any finite
$l$). This approximation is consistent with our weak coupling concept discussed
earlier and is valid for our gluonic application which entails
a nonperturbative confining potential in the diagonal sector Hamiltonian and a
weak, perturbative residual interaction in the off-diagonal.

We now solve the flow equations perturbatively, iterating $h_{pq}$.  We do not
treat $h_{pp'}$ perturbatively, but note it depends
very weakly upon $l$, being suppressed to second order in $h_{pq}$ since
>from Eqs.
(2.11,2.12),
$dE_{p}^{I}/dl\sim O(h_{pq}^2)$ and $dh_{pp'}^{I}/dl\sim O(h_{pq}^2)$.
Hence, our leading iteration in Eqs. (2.13,2.14) uses
$E_p^{I}(l)\sim E_p^{I}(0)=E_p(0)+\delta E_p^{I}(0)$
and $h_{pp'}^{I}(l)\sim h_{pp'}^{I}(0)$.  We could systematically iterate
Eqs. (2.11-2.14) carrying all terms, however, it is simpler and more direct
to further approximate Eqs. (2.13,2.14) by recognizing that the two
off-diagonal
sums over $p'$ and $q'$ will tend to be suppressed by cancellation.  To
appreciate
this point, consider the leading order iteration of Eq. (2.14) in Eq.
(2.13)  which
generates terms of the type,
$\sum_{p'}h_{pp'}^{I}(0)\eta_{p'q}(l)\sim\sum_{p'}
(E_{p'}^{I}(0)-E_q^{I}(0))h_{pp'}^{I}(0)h_{p'q}(l)$,
with a similar expression for the
$Q$-space sum, $\sum_{q'}\eta_{pq'}(l)h_{q'q}^{I}(0)$.  Here the sum is
over Fock
state components and all momentum (energy) for each Fock state.  Next we
make the
key observation that although the Q space has in general an infinite number
of Fock
states, the Fock sum over $q'$ is restricted to at most two adjacent
sectors since
$h_{qq'}$ is a two-body interaction which can change particle number at most by
two. Hence the $P$ and $Q$ space "volumes" in particle number (and energy for a
given particle number sector) are the same.  The two sums are therefore
equivalent to the product of an averaged matrix element for each space
with a common volume for both spaces.  Because the matrix elements have
random phases, in the limit of large numbers of energy states, the two
sums will tend to cancel and thus be suppressed relative to the leading
diagonal terms ($p = p', q = q'$).

Therefore in leading order the generator is
\begin{eqnarray}
\eta_{pq}(l) &=& -\frac{1}{E_p^{I}(0)-E_q^{I}(0)}\frac{dh_{pq}(l)}{dl}
\,,\label{eq:}\end{eqnarray}
and the effective Hamiltonian $PH_{eff}P$ is given by
$h^{eff}_{pp'} = h_{pp'}(\infty)$
\begin{eqnarray}
h^{eff}_{pp'} = E^I_p(0) \delta_{pp'}+h_{pp'}^{I}(0)|_{p\neq p'}
-\int_0^{\infty}dl\sum_{q}\left(\frac{\frac{dh_{pq}(l)}{dl}h_{qp'}(l)}{E_p^{I}(0
)-E_q^{I}(0)}+
\frac{h_{pq}(l)\frac{dh_{qp'}(l)}{dl}}{E^{I}_{p'}(0)-E_q^{I}(0)}\right)
\,,\label{eq:}\end{eqnarray}
having off-diagonal elements
\begin{eqnarray}
h_{pq}(l)=h_{pq}(0){\rm exp}\left(-l(E_p^{I}(0)-E_q^{I}(0))^2\right)
\,.\label{eq:}\end{eqnarray}
which can be used for the next iteration.
The above process can then be repeated,
replacing  $E^I_p(0)$, $h_{pp'}(0)$ on the right hand
side of the flow equations with the diagonal and off-diagonal matix elements of
$h^{eff}_{pp'}$, respectively, and continued until convergence.
Diagonalizing $h_{pp'}^{eff}$ then produces the final eigenvalues $E_p^{eff}$.
We now apply these
results to gluodynamics.

\section{Effective Hamiltonian for Gluodynamics}

In constructing our QCD motivated Hamiltonian, $H_{QCD}$, we have
incorporated  phenomenological elements of
previous, related constituent studies
\cite{ssjc96,flesrc}.  Starting with the exact
Coulomb gauge QCD Hamiltonian
(pure gluodynamics) \cite{zwanziger},
$\hat{H}_{QCD} = H_{free} + H^I_{QCD}$,
containing free and QCD interacting terms, we introduce a confining
phenomenological interaction, $H_{phen}$, to permit
isolation of the nonperturbative and perturbative Hamiltonians,
${ H}_{NP}$ and ${ H}_{PT}$, respectively
\begin{eqnarray}
{ \hat{H}}_{QCD}={ H}_{NP}+{ H}_{PT} \cong { H}_{NP}+
{H}^I_{can} \equiv { H}_{QCD}
\,,\label{eq:2.1}\end{eqnarray}
\begin{eqnarray}
{ H}_{NP}={ H}_{free}+{ H}_{phen}
\,,\label{eq:2.2}\end{eqnarray}
\begin{eqnarray}
{ H}_{PT}={ H}_{QCD}^{I}-{ H}_{phen} \cong { H}^I_{can}
\,.\label{eq:2.3}\end{eqnarray}
The soluable, phenomenological interaction ${ H}_{phen}$ is introduced
to provide a weaker, residual interaction ${ H}_{PT}$ at all
energy scales.  Following quark models, we assume that, in
the constituent (quasiparticle) basis, ${ H}_{NP}$ predominantly governs
the bound state bulk properties and should be treated
nonperturbatively. We further assume that the residual interaction can be
approximated at high energies by the canonical QCD interaction Hamiltonian
${ H}_{PT}={ H}_{can}^{I}(\Lambda_{UV}\rightarrow\infty)$
and is amenable to
perturbation theory. Note that the residual interaction describes
quantum fluctuations and may contribute UV divergent terms
requiring regularization with cut-off parameter $\Lambda_{UV}$.

The goal is to use the flow equations to scale the residual interaction
to low energies to generate an effective interaction, which
when added to ${ H}_{NP}$, can be diagonalized
nonperturbatively for bound states. In this process renormalization will
also be achieved by introducing second order
$O(g^2)$  canonical counterterms which remove the leading UV
divergences.
With an assumed weak residual interaction,
the entire procedure can be conducted perturbatively. There is no
over representation or double counting, since, as shown below, the
phenomenological interactions  determine the infrared (IR) behavior while the
residual interactions, which are free from IR divergences, describe
the UV part.

We now further detail the Hamiltonian.
In the Coulomb gauge the
physical degrees of freedom are the transverse gauge fields
${\bf A}^a T^a$ (with color matrices $T^a$ and sum over index $a$)
and their conjugate transverse momenta ${\bf\Pi}$.
The free Hamiltonian is
\begin{eqnarray}
{ H}_{free} = {\rm Tr}\int d {\bf x} \left( {\bf\Pi}^2({\bf x})
+ {\bf B}^2_{A}({\bf x}) \right)
\,,\label{eq:2.5}\end{eqnarray}
where ${\bf B}_{A} = \nabla \times \bf A$ is the abelian component
of the color magnetic field.  The nonabelian contribution,
${\bf B}^2({\bf x})-{\bf B}^2_{A}({\bf x})$,
is included in $H_{PT}$ (see below).  Here
${\bf B}={\bf B}^a T^a$ is the complete magnetic field having components
$B_i^a=\epsilon_{ijk}\nabla_jA_k^a+\frac{g}{2}\epsilon_{ijk}f^{abc}A_j^bA_k^c$,
with bare coupling constant $g$.
As in \cite{ssjc96} the phenomenological Hamiltonian is taken to be
\begin{eqnarray}
 { H}_{phen} =-\frac{1}{2}\int d{\bf x}d{\bf y}\rho^a({\bf x})V_L(|{\bf
x}-{\bf y}|)
\rho^a({\bf y})
\,,\label{eq:2.6}\end{eqnarray}
with color charge density (only gluon component)
$\rho^a({\bf x})=f^{abc}{\bf A}^b({\bf x}){\bf\Pi}^c({\bf x})$
and linear confining potential
\begin{eqnarray}
 V_L(|{\bf x}-{\bf y}|)=\sigma |{\bf x}-{\bf y}|
\,,\label{eq:2.7}\end{eqnarray}
with string tension, $\sigma=0.18\, GeV^2$, in accordance
with lattice calculations and Regge phenomenology.
The approximate residual interaction is given by the sum,
${ H}_{PT}={ H}_{NA} + { H}_C$,
with the nonabelian part
including three-, $H_{3g}$, and four-gluon, $H_{4g}$, interactions (see
Appendix A)
\begin{eqnarray}
{ H}_{NA} = {\rm Tr}\int d {\bf x}
              \left( {\bf B}^2({\bf x})-{\bf B}^2_{A}({\bf x}) \right)
             = { H}_{3g} + { H}_{4g}
\,,\label{eq:2.9}\end{eqnarray}
and the Coulomb interaction
\begin{eqnarray}
 { H}_C =-\frac{1}{2}\int d{\bf x}d{\bf y}\rho^a({\bf x})V_C(|{\bf x}-{\bf y}|)
\rho^a({\bf y})
\,,\label{eq:2.6a}\end{eqnarray}
\begin{eqnarray}
 V_C(|{\bf x}-{\bf y}|)=-C_{adj}\frac{\alpha_s}{|{\bf x}-{\bf y}|}
\,,\label{eq:2.8}\end{eqnarray}
having $\alpha_{s} = {g^2 \over 4 \pi}$ and adjoint Casimir operator
$C_{adj}=(N_c^2-1)/2N_c$ (here
$N_c = 3$ colors).
An advantage of the Coulomb gauge is the straightforward implemention
of phenomenological potentials. We define the instantaneous interaction,
${ H}_{L+C}$, by the combination
\begin{eqnarray}
{ H}_{L+C}={ H}_{phen}+{ H}_C
\,,\label{eq:2.9a}\end{eqnarray}
containing the sum, $V_L+V_C$,
of linear and Coulomb potentials.

Because of the phenomenological interaction, the trivial, perturbative vacuum
is not the minimum ground state for the QCD motivated Hamiltonian,
${ H}_{QCD}$.
Instead, we introduce
a trial nonperturbative vacuum state $|0\rangle$
which can be determined variationally, analogous to BSC type vacuum
\cite{ssjc96,flesrc,orsay,adler,brambilla} studies.
The Fock space is constructed
>from this vacuum using quasiparticle operators
$a_i^a({\bf k}), a_i^{a\dagger}(-{\bf k})$ which appear in the field
operator expansions
\begin{eqnarray}
 && A_i^a({\bf x}) = \int\frac{d{\bf
k}}{(2\pi)^3}\frac{1}{\sqrt{2\omega_{\bf k}}}
    [a_i^a({\bf k})+a_i^{a\dagger}(-{\bf k})]{\rm e}^{i{\bf k}{\bf x}}
\nonumber\\
 && \Pi_i^a({\bf x}) = -i\int\frac{d{\bf k}}{(2\pi)^3}\sqrt{
\frac{\omega_{\bf k}}{2} }
    [a_i^a({\bf k})-a_i^{a\dagger}(-{\bf k})]{\rm e}^{i{\bf k}{\bf x}}
\,,\label{eq:2.10}\end{eqnarray}
with $a_i^a({\bf k})|0\rangle=0$. Here $\omega_{\bf k}$
is the variational gluon energy from the gap equation (see below), obtained
by minimizing the ground state (vacuum) energy. The canonical commutation
relation is
\begin{eqnarray}
[a_i^a({\bf k}),a_j^{b\dagger}({\bf
k}^{\prime})]=(2\pi)^3\delta^{ab}\delta^{(3)}
({\bf k}-{\bf k}^{\prime})D_{ij}({\bf k})
\,,\label{eq:2.11}\end{eqnarray}
where the gluon operators
$a_i^a({\bf k})=\sum_{\lambda=1,2}\epsilon_i({\bf k},\lambda)a^a({\bf
k},\lambda)$
are transverse, i.e.
${\bf k}\cdot{\bf a}^a({\bf k})={\bf k}\cdot{\bf a}^{a\dagger}({\bf k})=0$,
and $D_{ij}({\bf k})$ is a polarization sum
\begin{eqnarray}
D_{ij}({\bf k})=\sum_{\lambda=1,2}
\epsilon_i({\bf k},\lambda)\epsilon_j({\bf k},\lambda)
=\delta_{ij}-{\hat k}_i{\hat k}_j
\,,\label{eq:2.12}\end{eqnarray}
with unit vector component ${\hat k}_i=k_i/k$ and $\hat k_i\cdot D_{ij}({\bf
k})=0$.

The complete QCD motivated Hamiltonian, Eq. (\ref{eq:2.1}), can now be
expressed in second quantized form and normal ordered
with respect to the trial vacuum.
In addition to two-body interactions, normal ordering leads to condensate
(vacuum
expectations) and one-body operators, both having  contributions from the
instantaneous and four-gluon terms.  The final expressions are
complicated and are summarized in Appendix A.

Finally the flow equations are applied to
${ H}_{QCD}={ H}_{free}+{ H}_{L+C}+{ H}_{3g}+{ H}_{4g}$.
Seperating the diagonal, ${ H}_d$, particle number conserving and
off-diagonal, ${ H}_r$, particle number changing parts yields
\begin{eqnarray}
{ H}_d &=& \left({ H}_{free}^{(0)}\right)_{nn} +
\left({ H}_{L+C}^{(0)+(2)}\right)_{nn} +
\left({ H}_{4g}^{(2)}\right)_{nn}
\,,\nonumber\\
{ H}_r &=&\left({ H}_{3g}^{(1)}\right)_{nm} +
\left({ H}_{free}^{(2)}\right)_{nm} +
\left({ H}_{C}^{(2)}\right)_{nm} +
\left({ H}_{4g}^{(2)}\right)_{nm}
\,,\label{eq:2.14}\end{eqnarray}
where $nn$  ($nm$) indicates particle number conserving (changing)
matrix elements. Here and throughout, the superscript
denotes the order (power) in
the bare coupling constant.  Note that the triple gluon Hamiltonian,
${H}_{3g}^{(1)}$, which is first order, only
contributes off-diagonally, while the four-gluon
Hamiltonian, ${ H}_{4g}^{(2)}$, which is second order,
contributes to both $H_d$ and $H_r$.
Also, as discussed previously, the linear confining potential, which is
zeroth order, is restricted to diagonal
contributions. All terms are detailed in Appendix A along with the free
Hamiltonian operator, containing the kinetic ($K, K^{(0)}$) and condensate
($O, O^{(0)}$) terms, which contributes
to the diagonal part in lowest order,
${ H}_{free}^{(0)}=K^{(0)}+O^{(0)}$, and to the off-diagonal part in
second order,
${ H}_{free}^{(2)}=(K-K^{(0)})+(O-O^{(0)})$.

Having identified the
diagonal and rest parts we now construct the generator,
$\eta$, and solve the flow
equations for the first two leading orders.
The leading order $O(g)$ (in coupling constant)
flow equation is
\begin{eqnarray}
\frac{dH_{3g}^{(1)}(l)}{dl} &=& [\eta^{(1)}(l), K^{(0)}(l)]\,,
\nonumber\\
\eta^{(1)}(l) &=& [K^{(0)}(l), H_{3g}^{(1)}(l)]
\,.\label{eq:2.15}\end{eqnarray}
Here the kinetic and triple-gluon vertex terms
are given (see Appendix A) respectively by
\begin{eqnarray}
K^{(0)}(l) &=& \int\frac{d{\bf k}}{(2\pi)^3}\omega ({\bf k}; l)
a_i^{a\dagger}({\bf k})a_i^a({\bf k})
\,,\label{eq:2.16}\end{eqnarray}
and in symmetrized form
\begin{eqnarray}
H_{3g}^{(1)}(l) &=& \frac{i}{2\sqrt{2}}f^{abc}
\int\left(\prod_{n=1}^{3} {d{\bf k}_n\over(2\pi)^3}\right)
(2\pi)^3\delta^{(3)}(\sum_m {\bf k}_m)
\frac{1}{\sqrt{\omega_1\omega_2\omega_3}}
\Gamma_{ijk}({\bf k}_1,{\bf k}_2,{\bf k}_3)
\nonumber \\
& &\hspace{-2cm}
\times {\bf :}
\left[g_0({\bf k}_1,{\bf k}_2,{\bf k}_3; l)
a^a_i({\bf k}_1) a^b_j({\bf k}_2) a^c_k({\bf k}_3)
+3 g_1({\bf k}_1,{\bf k}_2,{\bf k}_3; l)
a^{a\dagger}_i(-{\bf k}_1) a^b_j({\bf k}_2) a^c_k({\bf k}_3)
+{\rm h.c.} \right]
{\bf :}  
\,,\label{eq:2.16a}\end{eqnarray}
with shorthand notation $\omega_1=\omega({\bf k}_1)$, etc. and $\Gamma_{ijk}$
defined by
\begin{eqnarray}
\Gamma_{ijk}({\bf k}_1,{\bf k}_2,{\bf k}_3)=
\frac{1}{6}\left[
(k_1-k_3)_j\delta_{ik}+(k_2-k_1)_k\delta_{ij}+(k_3-k_2)_i\delta_{jk} \right]
\nonumber
\,,\label{eq:2.16b}\end{eqnarray}
satisfying
\begin{eqnarray}
&& \Gamma_{ijk}(-{\bf k}_1,-{\bf k}_2,-{\bf k}_3)=
-\Gamma_{ijk}({\bf k}_1,{\bf k}_2,{\bf k}_3)
\,,\nonumber\\
&& \Gamma_{jik}({\bf k}_2,{\bf k}_1,{\bf k}_3)=
-\Gamma_{ijk}({\bf k}_1,{\bf k}_2,{\bf k}_3)
\,.\label{eq:2.16c}\end{eqnarray}
Notice that in implementing the flow equations new, effective
coupling constants are generated (see below),
$g_{0}({\bf k}_1,{\bf k}_2,{\bf k}_3;l)$,
$g_{1}({\bf k}_1,{\bf k}_2,{\bf k}_3; l)$, which like the
energy, $\omega ({\bf k}; l)$, depend
upon the flow parameter $l$. The
coupling constants also aquire an effective energy/momentum dependence
corresponding to a given Fock sector
(different Fock sector operators flow differently in energy, indicated
by $0$ and $1$ subscripts here and below).
In leading order we omit the dependence of gluon energy on $l$ in the triple
vertex. The leading order generator, Eq. (\ref{eq:2.15}), is
\begin{eqnarray}
\eta^{(1)}(l) &=& \frac{i}{2\sqrt{2}}f^{abc}
\int\left(\prod_{n=1}^{3} {d{\bf k}_n\over(2\pi)^3}\right)
(2\pi)^3\delta^{(3)}(\sum_m {\bf k}_m)
\frac{1}{\sqrt{\omega_1\omega_2\omega_3}}
\Gamma_{ijk}({\bf k}_1,{\bf k}_2,{\bf k}_3) \label{eq:2.17} \\
& & \hspace{-2cm}
\times {\bf :}
\left[\eta_0({\bf k}_1,{\bf k}_2,{\bf k}_3;l)
a^a_i({\bf k}_1) a^b_j({\bf k}_2) a^c_k({\bf k}_3)
+3 \eta_1({\bf k}_1,{\bf k}_2,{\bf k}_3;l)
a^{a\dagger}_i(-{\bf k}_1) a^b_j({\bf k}_2) a^c_k({\bf k}_3)
-{\rm h.c.} \right]
{\bf :}  \nonumber
\,,\end{eqnarray}
where
\begin{eqnarray}
\eta_0({\bf k}_1,{\bf k}_2,{\bf k}_3;l) &=&
D_0(\omega_1,\omega_2,\omega_3)g_0({\bf k}_1,{\bf k}_2,{\bf k}_3;l)
\,,\nonumber\\
\eta_1({\bf k}_1,{\bf k}_2,{\bf k}_3;l) &=&
D_1(\omega_1,\omega_2,\omega_3)g_1({\bf k}_1,{\bf k}_2,{\bf k}_3;l)
\,,\label{eq:2.18}\end{eqnarray}
with energy terms
\begin{eqnarray}
D_0(\omega_1,\omega_2,\omega_3) = -(\omega_1+\omega_2+\omega_3)\,, \,
D_1(\omega_1,\omega_2,\omega_3) = -(-\omega_1+\omega_2+\omega_3)
\,.\label{eq:2.18a}\end{eqnarray}
The solutions of the flow equations, Eq.(3.15), for the
effective coupling constants are
\begin{eqnarray}
g_0(l)=g(0){\rm exp}(-D_0^2 l) \,, \,
g_1(l)=g(0){\rm exp}(-D_1^2 l)
\,,\label{eq:2.19}\end{eqnarray}
which, as anticipated, eliminates the triple-gluon vertex,
Eq. (\ref{eq:2.16a}),
for $l\rightarrow \infty$.
Correspondingly, new operators in the particle number conserving sectors
are generated, found from
the second order $O(g^2)$ flow equation
\begin{eqnarray}
\frac{dH_d^{(2)}(l)}{dl}=[\eta^{(1)}(l),H_{3g}^{(1)}(l)]
\,,\label{eq:2.20}\end{eqnarray}
which contribute to the effective two-body, self-energy and condensate
operators in the effective Hamiltonian.
\footnote{ One can always eliminate the particle number changing
terms, $H_r^{(2)}$, appearing in second order, by
choosing the generator $\eta^{(2)}=[K^{(0)}, H_r^{(2)}]$.
The flow equation reads
\begin{eqnarray}
\frac{dH_r^{(2)}}{dl}=[\eta^{(1)},H_{3g}^{(1)}]+[\eta^{(2)}, K^{(0)}]
\,,\label{eq:2.21}\end{eqnarray}
and, due to the second term in Eq. (\ref{eq:2.21}),
$H_r^{(2)}$ decays exponentially with flow parameter.
}
Solving the flow equations,
Eq. (\ref{eq:2.15}) and Eq. (\ref{eq:2.20}),
the block-diagonal effective Hamiltonian, $H_{ eff}$,
renormalized to the second order, is obtained
(details in Appendix B)
\begin{eqnarray}
{ H}_{eff} = { H}_d(\Lambda) + { H}_{gen}(\Lambda) +
\delta X_{CT}(\Lambda)
\,,\label{eq:2.21a}\end{eqnarray}
where ${ H}_d(\Lambda)$ is the particle number conserving part
of the original Hamiltonian,
Eq. (\ref{eq:2.14}), and ${ H}_{gen}(\Lambda)$
includes new operators generated via
perturbative elimination
of ${ H}_r$.  The cut-off parameter,
$\Lambda$, regulates UV divergences in loop integrals.
Renormalization is achieved through second order
by adding the mass counterterm, $\delta X_{CT}(\Lambda)$,
\begin{eqnarray}
\delta X_{CT}(\Lambda) &=& m_{CT}^2(\Lambda)
{\rm Tr}\int d{\bf x}{\bf A}^2({\bf x})
\,,\nonumber\\
m_{CT}^2(\Lambda) &=& -\frac{\alpha_s}{\pi}N_c\frac{11}{6}\Lambda^2
\,,\label{eq:2.21b}\end{eqnarray}
which absorbs the leading UV divergences in ${ H}_{eff}$
as $\Lambda\rightarrow \infty$. Therefore
we omit the cut-off notation in ${ H}_{eff}$, Eq. (\ref{eq:2.21a}),
because, through perturbative renormalization, the cut-off sensitivity is
weak (see
below).

We summarize the matrix elements of ${ H}_{eff}$
in the sectors of interest (up to two gluon states)
\footnote{Since calculations are done through second order
in the coupling constant,
new terms are generated only up to the two-body sector
(no three-body or higher interactions).}
\begin{eqnarray}
& & \langle 0|H_{ eff}|0\rangle =
 O^{ren} + O_{L+C}(\Lambda) + O_{4g}(\Lambda) + O_{\rm gen}(\Lambda)
\,,\nonumber\\
& & \langle 1|H_{ eff}|1\rangle =
 K^{ren} + \Pi_{L+C}(\Lambda) + \Pi_{4g}(\Lambda) + \Pi_{\rm gen}(\Lambda)
\,,\nonumber\\
& & \langle 2|H_{ eff}|2\rangle =
 V_{L+C} + V_{4g} + V_{\rm gen}
\,,\label{eq:2.22}\end{eqnarray}
where $|0\rangle$ is a shorthand notation for the zero-gluon sector
(also vacuum state), $|1\rangle$ the one-gluon sector, etc.
One should distinguish between two types of ${ H}_{eff}$ terms.
The first one arises from normal ordering the original Hamiltonian:
the instantaneous interaction with linear
plus Coulomb potentials, labeled by $L+C$, and the four-gluon vertex,
labeled by $4g$.
This leads to the polarization operators $\Pi_{L+C}$ and
$\Pi_{4g}$ in the one-body sector,
and condensate terms $O_{L+C}$ and $O_{4g}$ in the zero-body sector.
The energy of the ground state $O$ comes from normal ordering
the free Hamiltonian $H_{free}$ with respect to the vacuum $|0 \rangle$. As
shown
below,  given the phenomenological potential
$V_L$, the energy $O$ of the ground state  reproduces the known result
>from sum rules for the nonperturbative gluon condensate \cite{ShVaZa}.
The second type of terms are dynamical operators, generated by flow equations
and labeled $gen$.
In Eq. (\ref{eq:2.22}) the renormalized condensate $O^{ren}$
and kinetic $K^{ren}$ terms are
\begin{eqnarray}
O^{ren} &=& O + \delta X_{CT}^{0body}= \frac{1}{2}(N_c^2-1){\bf V}
\int\frac{d{\bf k}}{(2\pi)^3}\left( \frac{{\bf k}^2+m_{CT}^2(\Lambda)}
{\omega_{\bf k}}+\omega_{\bf k} \right)  \,,\nonumber\\
&=& (N_c^2-1){\bf V}
\int\frac{d{\bf k}}{(2\pi)^3}\omega_{\bf k}
+\frac{1}{2}(N_c^2-1){\bf V}
\int\frac{d{\bf k}}{(2\pi)^3}
\left( \frac{{\bf k}^2+m_{CT}^2(\Lambda)}
{\omega_{\bf k}} - \omega_{\bf k} \right)
\,,\\
K^{ren} &=& K + \delta X_{CT}^{1body}
=\int\frac{d{\bf k}}{(2\pi)^3}\omega_{\bf k}
a_i^{a\dagger}({\bf k})a_i^a({\bf k})
+\frac{1}{2}\int\frac{d{\bf k}}{(2\pi)^3}
\left( \frac{{\bf k}^2+m_{CT}^2(\Lambda)}
{\omega_{\bf k}} - \omega_{\bf k} \right)
\nonumber\\
& \times &
(a_i^{a\dagger}({\bf k})a_i^a({\bf k})
+\frac{1}{2}( a_i^a({\bf k})a_i^a(-{\bf k})+ {\rm h.c.} ))
\,,\label{eq:2.23}\end{eqnarray}
where $O$ and $K$ are defined by Eqs. (A1) and (A3), respectively,
and $\delta X_{CT}^{0body}$ is the mass counterterm,
given by Eq. (\ref{eq:2.21b})
in the zero-body sector (analogous for $\delta X_{CT}^{1body}$).
The mass counterterms $\delta X_{CT}$ cancel
the leading $\Lambda^2$ contribution from radiative corrections
to the vacuum and kinetic terms. The subleading
logarithmic dependence
remains, producing a slow cut-off dependence of eigenvalues
(see below).
In Eq. (\ref{eq:2.22}) the corrections to $O^{ren}$ and $K^{ren}$,
regulated by the exponential cut-off function, include the condensate terms
(Appendix B)
\begin{eqnarray}
O_{L+C}(\Lambda) &=& \frac{1}{16} N_c(N_c^2-1){\bf V}
\int{d{\bf k}\thinspace d{\bf q}\over(2\pi)^6}
{1\over\omega_{\bf k}\omega_{\bf q}}
\widetilde{V}_{L+C}({\bf k}-{\bf q})
(\omega_{\bf k}-\omega_{\bf q})^2
\left(1+({\hat k}{\hat q})^2\right)
{\rm e}^{-(|{\bf q}|+|{\bf k}|)^2/\Lambda^2}
\,,\nonumber\\
O_{4g}(\Lambda) &=& \frac{\alpha_s\pi}{4} N_c(N_c^2-1){\bf V}
\int{d{\bf k}\thinspace d{\bf q}\over(2\pi)^6}
{1\over\omega_{\bf k}\omega_{\bf q}}
\left(3-({\hat k}{\hat q})^2 \right)
{\rm e}^{-(|{\bf q}|+|{\bf k}|)^2/\Lambda^2}
\,,\label{eq:2.25}  \\
O_{\rm gen}(\Lambda) &=& -\alpha_s\pi N_c(N_c^2-1){\bf V}
\int {d{\bf k} d{\bf q} \over(2\pi)^6}
\frac{1}{3\omega_{\bf k}\omega_{\bf q}\omega_{{\bf k}-{\bf q}} }
\frac{G({\bf k},{\bf q})}{(\omega_{\bf k}+\omega_{\bf q}+\omega_{{\bf
k}-{\bf q}}) }
{\rm e}^{-(|{\bf q}|+|{\bf k}|+|{\bf k}-{\bf q}|)^2/\Lambda^2} \nonumber
\,,\end{eqnarray}
and the polarization operators (Appendix B)
\begin{eqnarray}
\Pi_{L+C}(\Lambda) &=& \frac{1}{8}N_c\int {d{\bf k} d{\bf q} \over(2\pi)^6}
\frac{1}{\omega_{\bf k}\omega_{\bf q}}
\widetilde{V}_{L+C}({\bf k}-{\bf q})(1+({\hat k}{\hat q})^2)
{\rm e}^{-q^2/\Lambda^2}
\nonumber\\
&\times & \left( (\omega_{\bf q}^2+\omega_{\bf k}^2)
a^{a\dagger}_i({\bf k})a^{a}_{i}({\bf k})
+(\omega_{\bf q}^2-\omega_{\bf k}^2)
\frac{1}{2}(a^{a}_i({\bf k})a^{a}_{i}({-\bf k})+{\rm h.c.}) \right)
\,,\nonumber \\
\Pi_{4g}(\Lambda) &=& \frac{\alpha_s\pi}{2}N_c
\int {d{\bf k} d{\bf q} \over(2\pi)^6}
\frac{1}{\omega_{\bf k}\omega_{\bf q}}(3-({\hat k}{\hat q})^2)
{\rm e}^{-q^2/\Lambda^2}
\nonumber\\
&\times & \left(a^{a\dagger}_i({\bf k})a^{a}_{i}({\bf k})
+\frac{1}{2}(a^{a}_i({\bf k})a^{a}_{i}({-\bf k})+{\rm h.c.}) \right)
\,,\nonumber\\
\Pi_{\rm gen}(\Lambda) &=& -\alpha_s\pi N_c
\int {d{\bf k} d{\bf q} \over(2\pi)^6}
\frac{1}{\omega_{\bf k}\omega_{\bf q}\omega_{{\bf k}-{\bf q}}}
\frac{G({\bf k},{\bf q})}{\omega_{\bf q}+\omega_{{\bf k}-{\bf q}} }
{\rm e}^{-4q^2/\Lambda^2}
\nonumber\\
&\times & \left(a^{a\dagger}_i({\bf k})a^{a}_{i}({\bf k})
+\frac{1}{2}(a^{a}_i({\bf k})a^{a}_{i}({-\bf k})+{\rm h.c.}) \right)
\,,\label{eq:2.6}\end{eqnarray}
containing a tensor structure in the generated term
\begin{eqnarray}
& & G({\bf k},{\bf q})=2(1-({\hat k}{\hat q})^2)
\left( k^2+q^2+\frac{k^2q^2}{2({\bf k}-{\bf q})^2}
(1+({\hat k}{\hat q})^2) \right)
\,,\label{eq:2.26a}\end{eqnarray}
and $\widetilde{V}_{L+C}$ is the momentum space representation of
the linear plus Coulomb potentials (see Appendix A).
Also, here and through out ${\hat k}{\hat q}= {\hat{\bf k}\cdot \hat {\bf
q}} $. The
effective gluon interaction in the color singlet channel is
\begin{eqnarray}
V_{eff} & &  = f^{abc}f^{a'b'c}
\int {d{\bf k} d{\bf q} \over(2\pi)^6}\frac{1}{\omega_{\bf k}\omega_{\bf q}}
V_{ii'jj'}({\bf k},{\bf q})
 \thinspace {\bf :}
{a^b_j}^\dagger({\bf q}){a^{b^\prime}_{j^\prime}}^\dagger(-{\bf q})
a^a_i({\bf k})a^{a^\prime}_{i^\prime}(-{\bf k})
{\bf :}
\,,\label{eq:2.27}\end{eqnarray}
which describes the bound state of a glueball at rest.
It includes the three interactions, $V_{L+C}+V_{4g}+V_{gen}$,
that define the effective Hamiltonian
in the two-body sector $\langle 2|H_{ eff}|2\rangle$,
Eq. (\ref{eq:2.22}).
\footnote{We do not consider color octet states.}
Explicitly (see Appendix B)
\begin{eqnarray}
  V^{L+C}_{ii'jj'} &=& -\frac{1}{8}\widetilde{V}_{L+C}({\bf k}-{\bf q})
(\omega_{\bf k}+\omega_{\bf q})^2 \delta_{ij}\delta_{i'j'}
\,,\nonumber\\
  V^{4g}_{ii'jj'} &=& \frac{\alpha_s\pi}{2}
(\delta_{ii'}\delta_{jj'}-\delta_{ij'}\delta_{i'j})
\,,\nonumber\\
V^{\rm gen}_{ii'jj'} &=& - \alpha_s 2\pi
\frac{1}{\omega_{{\bf k}-{\bf q}}^2}
T_{ij,i'j'}({\bf k},{\bf q})
\left(1-\frac{(\omega_{\bf k}-\omega_{\bf q})^2}
{(\omega_{\bf k}-\omega_{\bf q})^2+\omega_{{\bf k}-{\bf q}}^2}
\right)
\,,\label{eq:2.28}\end{eqnarray}
where the tensor term in the generated interaction is
\begin{eqnarray}
T_{ij,i'j'}({\bf k},{\bf q}) &=& ({\bf k}-{\bf q})^2
\left( \phantom{\frac{1}{2}}\hspace{-0.3cm}
n_in_{i'}\delta_{jj'}+n_jn_{j'}\delta_{ii'}
+n_in_j\delta_{i'j'}+n_{i'}n_{j'}\delta_{ij}
-n_in_{j'}\delta_{i'j}-n_{i'}n_j\delta_{ij'} \right.
\nonumber\\
&+& \left.\frac{k^2q^2}{({\bf k}-{\bf q})^4}(1-({\hat k}{\hat q})^2)
\delta_{ij}\delta_{i'j'}  \right)
\,,\label{eq:2.29}\end{eqnarray}
and $n_i$ are the components of the unit vector ${\bf n}=({\bf k}-{\bf
q})/|{\bf k}-{\bf q}|$. All other possible two-body interactions that change
particle number can be eliminated by the flow equations through second order
and only contribute in higher orders.
In the next section we utilize this effective, gluodynamical Hamiltonian,
with matrix elements given by Eq. (\ref{eq:2.22}),
to obtain and solve the gluon gap and
glueball bound state equations.

\section{Gap Equation and Glueballs}

Now that we have the block diagonal, effective Hamiltonian, ${ H}_{eff}$,
valid through second order in the coupling constant, we can
nonperturbatively diagonalize the $PH_{eff}P$ sector Hamiltonian.
Further, because ${ H}_{eff}$ is also renormalized,
equations for physical observables
are free from the leading UV divergences.  In subsection A we first investigate
the vacuum by formulating and solving the gluon gap equation and
also calculate the gluon condensate.  Then we address the glueball spectrum
in subsection B.

\subsection{Gap Equation}

The gap equation allows determination of a nontrivial vacuum with
gluon condensates and propagating quasiparticles,
here gluons with a  dynamical
mass. There are several ways to obtain this equation, the most common
based upon a variational principle to  minimize the vacuum (ground state)
energy. In the Bogoliubov-Valatin or BCS approach, the variational parameter
is  the angle of transformation from undressed to dressed particle
(quasiparticle)
operators.
In this work the variational parameter is the quasiparticle energy,
$\omega_{\bf k}$, which defines a quasiparticle basis,
Eq. (\ref{eq:2.10}),
where the effective Hamiltonian is block-diagonal.
Accordingly, minimizing
the vacuum energy of the effective Hamiltonian
\begin{eqnarray}
\frac{\delta \langle 0| {H}_{ eff}|0\rangle}{\delta\omega_{\bf k}}=0
\,,\label{eq:3.1}\end{eqnarray}
generates the gap equation for the unknown $\omega_{\bf k}$.
Using Eq. (\ref{eq:2.25}) for the condensate terms
the following gap equation is obtained
\begin{eqnarray}
\omega_{\bf k}^2 = k^2+m_{CT}^2(\Lambda)
+ \frac{1}{4} N_c
\int{d{\bf q}\over(2\pi)^3}{1\over\omega_{\bf q}}
\widetilde{V}_{L+C}({\bf k}-{\bf q})\left(1+({\hat k}{\hat q})^2 \right)
(\omega_{\bf q}^2-\omega_{\bf k}^2){\rm e}^{-q^2/\Lambda^2}
\nonumber\\
+ \alpha_s\pi N_c
\int{d{\bf q}\over(2\pi)^3}{1\over\omega_{\bf q}}
\left(3-({\hat k}{\hat q})^2 \right)
{\rm e}^{-q^2/\Lambda^2}
- 2\alpha_s\pi N_c
\int{d{\bf q}\over(2\pi)^3}{1\over\omega_{\bf q}\omega_{{\bf k}-{\bf q}} }
\frac{G({\bf k},{\bf q})}{\omega_{\bf q}+\omega_{{\bf k}-{\bf q}}}
{\rm e}^{-4q^2/\Lambda^2}
\,,\label{eq:3.2}\end{eqnarray}
where the mass counterterm $m_{CT}(\Lambda)$ is given by
Eq. (\ref{eq:2.21b}) and $G({\bf k},{\bf q})$ by
Eq. (\ref{eq:2.26a}).  Note that a
momentum regulating function has been introduced in the first two
integrals, while the third is naturally regulated from the
flow equations.

The gap equation can also be obtained by using
a renormalization group condition,
namely the invariance of Hamiltonian form
under the renormalization group transformation.
Since flow equations are in this class of
transformations,
the renormalized effective Hamiltonian
should not contain off-diagonal one-body terms of the type
$a({\bf k})a(-{\bf k})$ or
$a^{\dagger}({\bf k})a^{\dagger}(-{\bf k})$.
Imposing this condition on the matrix elements of the effective Hamiltonian
in the one-gluon sector $\langle 1|H_{ eff}|1\rangle$, Eq. (3.27)
and Eq. (\ref{eq:2.6}),
yields the same gap equation
as above.

There are two types of UV divergences
in the gap equation:
the leading, $\Lambda^2$, and the subleading, $\ln \Lambda$,
corresponding to relevant and marginal operators
in the context of the renormalization group, respectively.
Terms from the canonical Hamiltonian ${ H}_{PT}$
generate both types, while the confining interaction
>from ${ H}_{NP}$  only generates a $\ln \Lambda$ dependence.
Renormalization is acheived by finding the counterterms
associated with the canonical terms,
since only the canonical part is renormalizable.
We do not renormalize the confining interaction as this would
only lead to
"noncanonical" divergences in each order of perturbation theory.
The mass counterterm,
Eq. (\ref{eq:2.21b}), cancels the leading UV divergency
in the gap equation, Eq. (\ref{eq:3.2}), while
the uncanceled part
leads to a gap energy with a logarithmic cut-off dependence,
$\omega({\bf {k}}, \Lambda)$.

\begin{figure}[!htb]
\begin{center}
\input{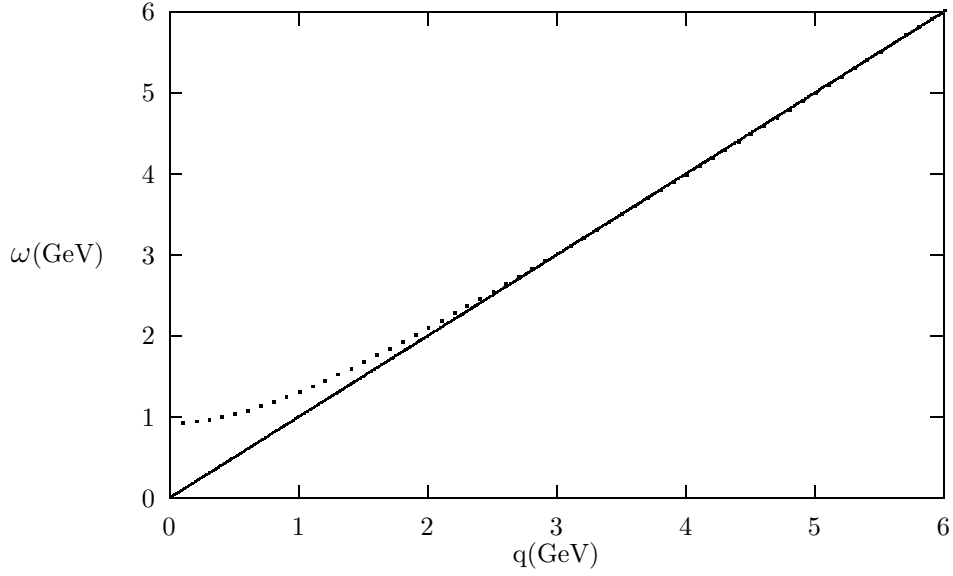}
\end{center}
\caption{One particle dispersion relation, $\omega({\bf q})$.
Dots are the numerical solution
of the gap equation for parameters
$\alpha_s=0.4, \, \sigma=0.18 \, GeV^2$ and $\Lambda=4 \, GeV$.
The solid line is the free
dispersion, $\omega({\bf q})=q$.}
\label{fig.1}
\end{figure}

\begin{figure}[!htb]
\begin{center}
\input{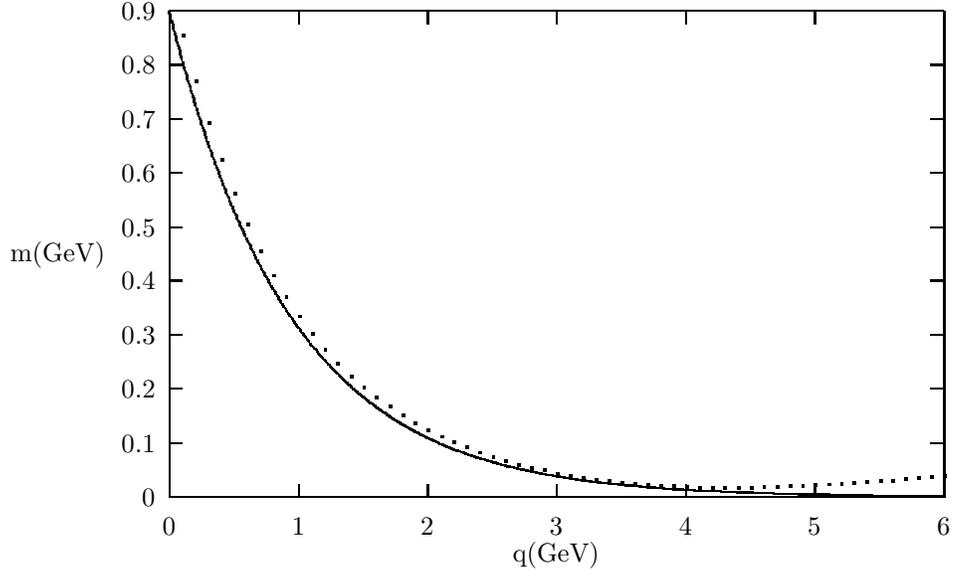}
\end{center}
\caption{Dynamical gluon mass. Dots represent the numerical solution
for $m({\bf q})=\omega({\bf q})-q$
(same parameters as in Fig. 1).
Solid line is
$m({\bf q})=0.9{\rm exp}(-q/0.95)$ (parameters are in GeV).}
\label{fig.2}
\end{figure}

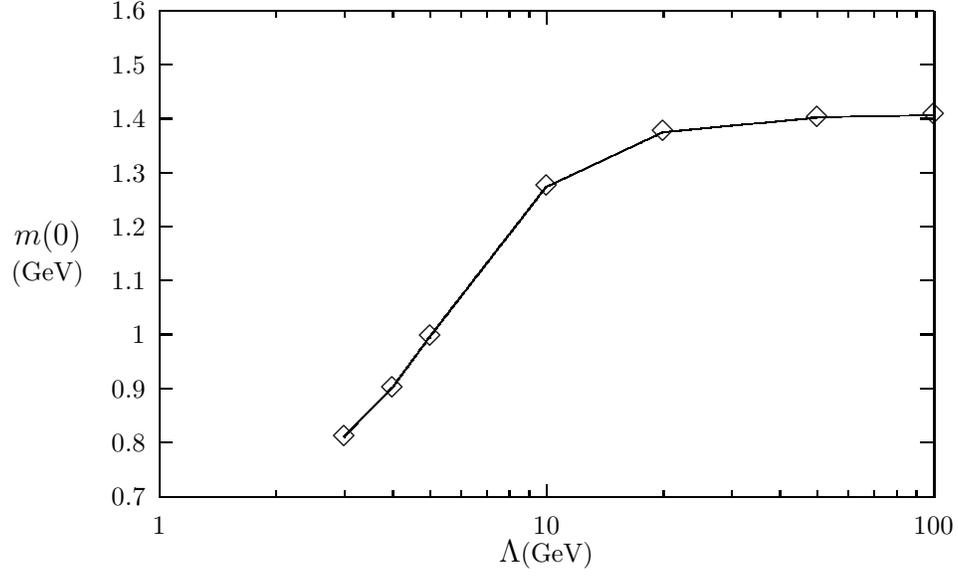
\begin{figure}[!htb]
\begin{center}
\setlength{\unitlength}{0.240900pt}
\ifx\plotpoint\undefined\newsavebox{\plotpoint}\fi
\sbox{\plotpoint}{\rule[-0.200pt]{0.400pt}{0.400pt}}%
\begin{picture}(1500,900)(0,0)
\font\gnuplot=cmr10 at 10pt
\gnuplot
\sbox{\plotpoint}{\rule[-0.200pt]{0.400pt}{0.400pt}}%
\put(220.0,113.0){\rule[-0.200pt]{4.818pt}{0.400pt}}
\put(198,113){\makebox(0,0)[r]{0.7}}
\put(1416.0,113.0){\rule[-0.200pt]{4.818pt}{0.400pt}}
\put(220.0,198.0){\rule[-0.200pt]{4.818pt}{0.400pt}}
\put(198,198){\makebox(0,0)[r]{0.8}}
\put(1416.0,198.0){\rule[-0.200pt]{4.818pt}{0.400pt}}
\put(220.0,283.0){\rule[-0.200pt]{4.818pt}{0.400pt}}
\put(198,283){\makebox(0,0)[r]{0.9}}
\put(1416.0,283.0){\rule[-0.200pt]{4.818pt}{0.400pt}}
\put(220.0,368.0){\rule[-0.200pt]{4.818pt}{0.400pt}}
\put(198,368){\makebox(0,0)[r]{1}}
\put(1416.0,368.0){\rule[-0.200pt]{4.818pt}{0.400pt}}
\put(220.0,453.0){\rule[-0.200pt]{4.818pt}{0.400pt}}
\put(198,453){\makebox(0,0)[r]{1.1}}
\put(1416.0,453.0){\rule[-0.200pt]{4.818pt}{0.400pt}}
\put(220.0,537.0){\rule[-0.200pt]{4.818pt}{0.400pt}}
\put(198,537){\makebox(0,0)[r]{1.2}}
\put(1416.0,537.0){\rule[-0.200pt]{4.818pt}{0.400pt}}
\put(220.0,622.0){\rule[-0.200pt]{4.818pt}{0.400pt}}
\put(198,622){\makebox(0,0)[r]{1.3}}
\put(1416.0,622.0){\rule[-0.200pt]{4.818pt}{0.400pt}}
\put(220.0,707.0){\rule[-0.200pt]{4.818pt}{0.400pt}}
\put(198,707){\makebox(0,0)[r]{1.4}}
\put(1416.0,707.0){\rule[-0.200pt]{4.818pt}{0.400pt}}
\put(220.0,792.0){\rule[-0.200pt]{4.818pt}{0.400pt}}
\put(198,792){\makebox(0,0)[r]{1.5}}
\put(1416.0,792.0){\rule[-0.200pt]{4.818pt}{0.400pt}}
\put(220.0,877.0){\rule[-0.200pt]{4.818pt}{0.400pt}}
\put(198,877){\makebox(0,0)[r]{1.6}}
\put(1416.0,877.0){\rule[-0.200pt]{4.818pt}{0.400pt}}
\put(220.0,113.0){\rule[-0.200pt]{0.400pt}{4.818pt}}
\put(220,68){\makebox(0,0){1}}
\put(220.0,857.0){\rule[-0.200pt]{0.400pt}{4.818pt}}
\put(403.0,113.0){\rule[-0.200pt]{0.400pt}{2.409pt}}
\put(403.0,867.0){\rule[-0.200pt]{0.400pt}{2.409pt}}
\put(510.0,113.0){\rule[-0.200pt]{0.400pt}{2.409pt}}
\put(510.0,867.0){\rule[-0.200pt]{0.400pt}{2.409pt}}
\put(586.0,113.0){\rule[-0.200pt]{0.400pt}{2.409pt}}
\put(586.0,867.0){\rule[-0.200pt]{0.400pt}{2.409pt}}
\put(645.0,113.0){\rule[-0.200pt]{0.400pt}{2.409pt}}
\put(645.0,867.0){\rule[-0.200pt]{0.400pt}{2.409pt}}
\put(693.0,113.0){\rule[-0.200pt]{0.400pt}{2.409pt}}
\put(693.0,867.0){\rule[-0.200pt]{0.400pt}{2.409pt}}
\put(734.0,113.0){\rule[-0.200pt]{0.400pt}{2.409pt}}
\put(734.0,867.0){\rule[-0.200pt]{0.400pt}{2.409pt}}
\put(769.0,113.0){\rule[-0.200pt]{0.400pt}{2.409pt}}
\put(769.0,867.0){\rule[-0.200pt]{0.400pt}{2.409pt}}
\put(800.0,113.0){\rule[-0.200pt]{0.400pt}{2.409pt}}
\put(800.0,867.0){\rule[-0.200pt]{0.400pt}{2.409pt}}
\put(828.0,113.0){\rule[-0.200pt]{0.400pt}{4.818pt}}
\put(828,68){\makebox(0,0){10}}
\put(828.0,857.0){\rule[-0.200pt]{0.400pt}{4.818pt}}
\put(1011.0,113.0){\rule[-0.200pt]{0.400pt}{2.409pt}}
\put(1011.0,867.0){\rule[-0.200pt]{0.400pt}{2.409pt}}
\put(1118.0,113.0){\rule[-0.200pt]{0.400pt}{2.409pt}}
\put(1118.0,867.0){\rule[-0.200pt]{0.400pt}{2.409pt}}
\put(1194.0,113.0){\rule[-0.200pt]{0.400pt}{2.409pt}}
\put(1194.0,867.0){\rule[-0.200pt]{0.400pt}{2.409pt}}
\put(1253.0,113.0){\rule[-0.200pt]{0.400pt}{2.409pt}}
\put(1253.0,867.0){\rule[-0.200pt]{0.400pt}{2.409pt}}
\put(1301.0,113.0){\rule[-0.200pt]{0.400pt}{2.409pt}}
\put(1301.0,867.0){\rule[-0.200pt]{0.400pt}{2.409pt}}
\put(1342.0,113.0){\rule[-0.200pt]{0.400pt}{2.409pt}}
\put(1342.0,867.0){\rule[-0.200pt]{0.400pt}{2.409pt}}
\put(1377.0,113.0){\rule[-0.200pt]{0.400pt}{2.409pt}}
\put(1377.0,867.0){\rule[-0.200pt]{0.400pt}{2.409pt}}
\put(1408.0,113.0){\rule[-0.200pt]{0.400pt}{2.409pt}}
\put(1408.0,867.0){\rule[-0.200pt]{0.400pt}{2.409pt}}
\put(1436.0,113.0){\rule[-0.200pt]{0.400pt}{4.818pt}}
\put(1436,68){\makebox(0,0){100}}
\put(1436.0,857.0){\rule[-0.200pt]{0.400pt}{4.818pt}}
\put(220.0,113.0){\rule[-0.200pt]{292.934pt}{0.400pt}}
\put(1436.0,113.0){\rule[-0.200pt]{0.400pt}{184.048pt}}
\put(220.0,877.0){\rule[-0.200pt]{292.934pt}{0.400pt}}
\put(45,495){\makebox(0,0){\shortstack{$m(0)$\\(GeV)}}}
\put(828,23){\makebox(0,0){$\Lambda$(GeV)}}
\put(220.0,113.0){\rule[-0.200pt]{0.400pt}{184.048pt}}
\put(510,207){\usebox{\plotpoint}}
\multiput(510.58,207.00)(0.499,0.506){149}{\rule{0.120pt}{0.505pt}}
\multiput(509.17,207.00)(76.000,75.951){2}{\rule{0.400pt}{0.253pt}}
\multiput(586.58,284.00)(0.499,0.678){115}{\rule{0.120pt}{0.642pt}}
\multiput(585.17,284.00)(59.000,78.667){2}{\rule{0.400pt}{0.321pt}}
\multiput(645.58,364.00)(0.500,0.645){363}{\rule{0.120pt}{0.616pt}}
\multiput(644.17,364.00)(183.000,234.722){2}{\rule{0.400pt}{0.308pt}}
\multiput(828.00,600.58)(1.066,0.499){169}{\rule{0.951pt}{0.120pt}}
\multiput(828.00,599.17)(181.026,86.000){2}{\rule{0.476pt}{0.400pt}}
\multiput(1011.00,686.58)(5.336,0.496){43}{\rule{4.309pt}{0.120pt}}
\multiput(1011.00,685.17)(233.057,23.000){2}{\rule{2.154pt}{0.400pt}}
\multiput(1253.00,709.60)(26.655,0.468){5}{\rule{18.400pt}{0.113pt}}
\multiput(1253.00,708.17)(144.810,4.000){2}{\rule{9.200pt}{0.400pt}}
\put(510,207){\raisebox{-.8pt}{\makebox(0,0){$\Diamond$}}}
\put(586,284){\raisebox{-.8pt}{\makebox(0,0){$\Diamond$}}}
\put(645,364){\raisebox{-.8pt}{\makebox(0,0){$\Diamond$}}}
\put(828,600){\raisebox{-.8pt}{\makebox(0,0){$\Diamond$}}}
\put(1011,686){\raisebox{-.8pt}{\makebox(0,0){$\Diamond$}}}
\put(1253,709){\raisebox{-.8pt}{\makebox(0,0){$\Diamond$}}}
\put(1436,713){\raisebox{-.8pt}{\makebox(0,0){$\Diamond$}}}
\end{picture}
\end{center}
\caption{Cut-off dependence of the constituent gluon mass
(same parameters as in Fig. 1).}
\label{fig.3}
\end{figure}

We have numerically solved the linearized gap equation retaining only the
dominant,
instantaneous term, with Coulomb and confining potentials, which leads to a
different
corresponding Coulomb mass counterterm
\begin{eqnarray}
\widetilde{m}_{CT}^2(\Lambda) = -\frac{\alpha_s}{\pi}\frac{2N_c}{3}\Lambda^2
\,,\label{eq:3.3}\end{eqnarray}
instead of $m_{CT}^2$ which is only appropriate for
the full gap equation.
For fixed cut-off $\Lambda = 4 \, GeV$,  the energy dispersion,
$\omega({\bf k})$, is displayed
in Fig. (\ref{fig.1}). The free behavior, $\omega({\bf k})=k$,
is recovered at high energies while for low energies a constituent gluon
mass, roughly .9 $GeV$ is obtained.
As shown in Fig. (\ref{fig.2}), the analytic form,
$\omega({\bf k})=k + m({\bf k})$,
with a dynamical gluon mass, $m({\bf k})$, given by
\begin{eqnarray}
m({\bf k})=m(0){\rm exp}(-\frac{k}{\kappa})
\,,\label{eq:3.4}\end{eqnarray}
reproduces the numerical solution using parameters $m(0)=0.9$ $GeV$ and
$\kappa=0.95$
$GeV$. The sensitivity of the effective gluon mass at zero momentum, $m(0)$, to
cut-off $\Lambda$ is displayed in Fig. (\ref{fig.3}).

\begin{figure}[!htb]
\begin{center}
\setlength{\unitlength}{0.240900pt}
\ifx\plotpoint\undefined\newsavebox{\plotpoint}\fi
\begin{picture}(1500,900)(0,0)
\font\gnuplot=cmr10 at 10pt
\gnuplot
\sbox{\plotpoint}{\rule[-0.200pt]{0.400pt}{0.400pt}}%
\put(220.0,113.0){\rule[-0.200pt]{4.818pt}{0.400pt}}
\put(198,113){\makebox(0,0)[r]{0.01}}
\put(1416.0,113.0){\rule[-0.200pt]{4.818pt}{0.400pt}}
\put(220.0,189.0){\rule[-0.200pt]{4.818pt}{0.400pt}}
\put(198,189){\makebox(0,0)[r]{0.011}}
\put(1416.0,189.0){\rule[-0.200pt]{4.818pt}{0.400pt}}
\put(220.0,266.0){\rule[-0.200pt]{4.818pt}{0.400pt}}
\put(198,266){\makebox(0,0)[r]{0.012}}
\put(1416.0,266.0){\rule[-0.200pt]{4.818pt}{0.400pt}}
\put(220.0,342.0){\rule[-0.200pt]{4.818pt}{0.400pt}}
\put(198,342){\makebox(0,0)[r]{0.013}}
\put(1416.0,342.0){\rule[-0.200pt]{4.818pt}{0.400pt}}
\put(220.0,419.0){\rule[-0.200pt]{4.818pt}{0.400pt}}
\put(198,419){\makebox(0,0)[r]{0.014}}
\put(1416.0,419.0){\rule[-0.200pt]{4.818pt}{0.400pt}}
\put(220.0,495.0){\rule[-0.200pt]{4.818pt}{0.400pt}}
\put(198,495){\makebox(0,0)[r]{0.015}}
\put(1416.0,495.0){\rule[-0.200pt]{4.818pt}{0.400pt}}
\put(220.0,571.0){\rule[-0.200pt]{4.818pt}{0.400pt}}
\put(198,571){\makebox(0,0)[r]{0.016}}
\put(1416.0,571.0){\rule[-0.200pt]{4.818pt}{0.400pt}}
\put(220.0,648.0){\rule[-0.200pt]{4.818pt}{0.400pt}}
\put(198,648){\makebox(0,0)[r]{0.017}}
\put(1416.0,648.0){\rule[-0.200pt]{4.818pt}{0.400pt}}
\put(220.0,724.0){\rule[-0.200pt]{4.818pt}{0.400pt}}
\put(198,724){\makebox(0,0)[r]{0.018}}
\put(1416.0,724.0){\rule[-0.200pt]{4.818pt}{0.400pt}}
\put(220.0,801.0){\rule[-0.200pt]{4.818pt}{0.400pt}}
\put(198,801){\makebox(0,0)[r]{0.019}}
\put(1416.0,801.0){\rule[-0.200pt]{4.818pt}{0.400pt}}
\put(220.0,877.0){\rule[-0.200pt]{4.818pt}{0.400pt}}
\put(198,877){\makebox(0,0)[r]{0.02}}
\put(1416.0,877.0){\rule[-0.200pt]{4.818pt}{0.400pt}}
\put(220.0,113.0){\rule[-0.200pt]{0.400pt}{4.818pt}}
\put(220,68){\makebox(0,0){1}}
\put(220.0,857.0){\rule[-0.200pt]{0.400pt}{4.818pt}}
\put(403.0,113.0){\rule[-0.200pt]{0.400pt}{2.409pt}}
\put(403.0,867.0){\rule[-0.200pt]{0.400pt}{2.409pt}}
\put(510.0,113.0){\rule[-0.200pt]{0.400pt}{2.409pt}}
\put(510.0,867.0){\rule[-0.200pt]{0.400pt}{2.409pt}}
\put(586.0,113.0){\rule[-0.200pt]{0.400pt}{2.409pt}}
\put(586.0,867.0){\rule[-0.200pt]{0.400pt}{2.409pt}}
\put(645.0,113.0){\rule[-0.200pt]{0.400pt}{2.409pt}}
\put(645.0,867.0){\rule[-0.200pt]{0.400pt}{2.409pt}}
\put(693.0,113.0){\rule[-0.200pt]{0.400pt}{2.409pt}}
\put(693.0,867.0){\rule[-0.200pt]{0.400pt}{2.409pt}}
\put(734.0,113.0){\rule[-0.200pt]{0.400pt}{2.409pt}}
\put(734.0,867.0){\rule[-0.200pt]{0.400pt}{2.409pt}}
\put(769.0,113.0){\rule[-0.200pt]{0.400pt}{2.409pt}}
\put(769.0,867.0){\rule[-0.200pt]{0.400pt}{2.409pt}}
\put(800.0,113.0){\rule[-0.200pt]{0.400pt}{2.409pt}}
\put(800.0,867.0){\rule[-0.200pt]{0.400pt}{2.409pt}}
\put(828.0,113.0){\rule[-0.200pt]{0.400pt}{4.818pt}}
\put(828,68){\makebox(0,0){10}}
\put(828.0,857.0){\rule[-0.200pt]{0.400pt}{4.818pt}}
\put(1011.0,113.0){\rule[-0.200pt]{0.400pt}{2.409pt}}
\put(1011.0,867.0){\rule[-0.200pt]{0.400pt}{2.409pt}}
\put(1118.0,113.0){\rule[-0.200pt]{0.400pt}{2.409pt}}
\put(1118.0,867.0){\rule[-0.200pt]{0.400pt}{2.409pt}}
\put(1194.0,113.0){\rule[-0.200pt]{0.400pt}{2.409pt}}
\put(1194.0,867.0){\rule[-0.200pt]{0.400pt}{2.409pt}}
\put(1253.0,113.0){\rule[-0.200pt]{0.400pt}{2.409pt}}
\put(1253.0,867.0){\rule[-0.200pt]{0.400pt}{2.409pt}}
\put(1301.0,113.0){\rule[-0.200pt]{0.400pt}{2.409pt}}
\put(1301.0,867.0){\rule[-0.200pt]{0.400pt}{2.409pt}}
\put(1342.0,113.0){\rule[-0.200pt]{0.400pt}{2.409pt}}
\put(1342.0,867.0){\rule[-0.200pt]{0.400pt}{2.409pt}}
\put(1377.0,113.0){\rule[-0.200pt]{0.400pt}{2.409pt}}
\put(1377.0,867.0){\rule[-0.200pt]{0.400pt}{2.409pt}}
\put(1408.0,113.0){\rule[-0.200pt]{0.400pt}{2.409pt}}
\put(1408.0,867.0){\rule[-0.200pt]{0.400pt}{2.409pt}}
\put(1436.0,113.0){\rule[-0.200pt]{0.400pt}{4.818pt}}
\put(1436,68){\makebox(0,0){100}}
\put(1436.0,857.0){\rule[-0.200pt]{0.400pt}{4.818pt}}
\put(220.0,113.0){\rule[-0.200pt]{292.934pt}{0.400pt}}
\put(1436.0,113.0){\rule[-0.200pt]{0.400pt}{184.048pt}}
\put(220.0,877.0){\rule[-0.200pt]{292.934pt}{0.400pt}}
\put(45,495){\makebox(0,0){\shortstack{\hspace{-1cm}$\langle
F_{\mu\nu}F_{\mu\nu} \rangle$\hspace{-0.4cm}\\ \hspace{-1cm}$(GeV^4)$}}}
\put(828,23){\makebox(0,0){$\Lambda$(GeV)}}
\put(220.0,113.0){\rule[-0.200pt]{0.400pt}{184.048pt}}
\put(403,148){\usebox{\plotpoint}}
\multiput(403.58,148.00)(0.499,0.725){211}{\rule{0.120pt}{0.679pt}}
\multiput(402.17,148.00)(107.000,153.590){2}{\rule{0.400pt}{0.340pt}}
\multiput(510.00,303.58)(0.527,0.499){141}{\rule{0.522pt}{0.120pt}}
\multiput(510.00,302.17)(74.916,72.000){2}{\rule{0.261pt}{0.400pt}}
\multiput(586.00,375.58)(0.758,0.498){75}{\rule{0.705pt}{0.120pt}}
\multiput(586.00,374.17)(57.536,39.000){2}{\rule{0.353pt}{0.400pt}}
\multiput(645.00,414.58)(1.611,0.499){111}{\rule{1.384pt}{0.120pt}}
\multiput(645.00,413.17)(180.127,57.000){2}{\rule{0.692pt}{0.400pt}}
\multiput(828.00,471.58)(6.243,0.494){27}{\rule{4.980pt}{0.119pt}}
\multiput(828.00,470.17)(172.664,15.000){2}{\rule{2.490pt}{0.400pt}}
\multiput(1011.00,486.59)(26.870,0.477){7}{\rule{19.460pt}{0.115pt}}
\multiput(1011.00,485.17)(201.610,5.000){2}{\rule{9.730pt}{0.400pt}}
\put(403,148){\raisebox{-.8pt}{\makebox(0,0){$\Diamond$}}}
\put(510,303){\raisebox{-.8pt}{\makebox(0,0){$\Diamond$}}}
\put(586,375){\raisebox{-.8pt}{\makebox(0,0){$\Diamond$}}}
\put(645,414){\raisebox{-.8pt}{\makebox(0,0){$\Diamond$}}}
\put(828,471){\raisebox{-.8pt}{\makebox(0,0){$\Diamond$}}}
\put(1011,486){\raisebox{-.8pt}{\makebox(0,0){$\Diamond$}}}
\put(1253,491){\raisebox{-.8pt}{\makebox(0,0){$\Diamond$}}}
\put(1436,491){\raisebox{-.8pt}{\makebox(0,0){$\Diamond$}}}
\put(1253.0,491.0){\rule[-0.200pt]{44.085pt}{0.400pt}}
\end{picture}
\end{center}
\caption{Gluon condensate cut-off dependence
(same parameters as in Fig. 1).}
\label{fig.4}
\end{figure}
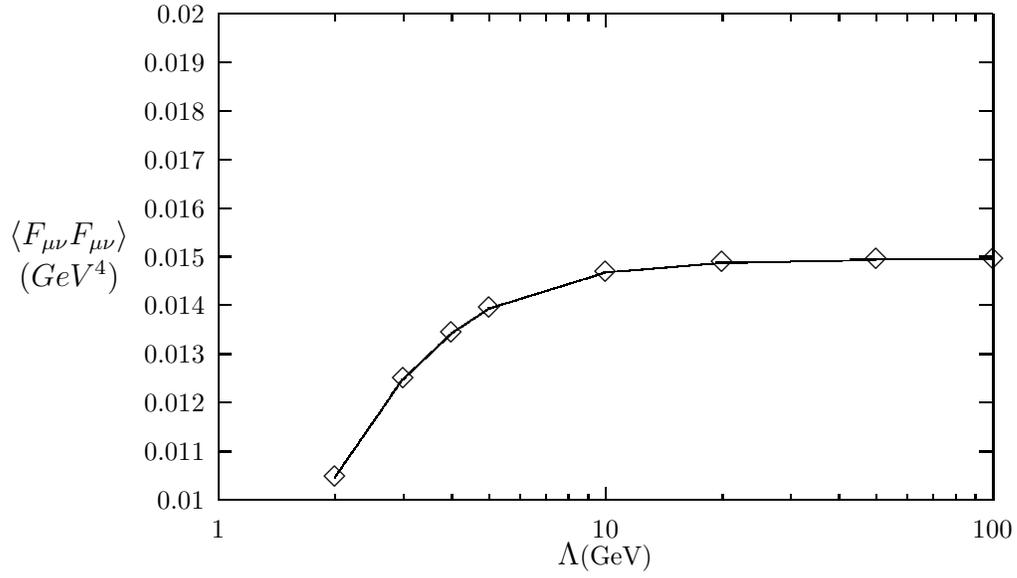

The gluon condensate is another nonperturbative quantity calculated
within this approach. The condensate is given by the vacuum expectation value
$O = \langle 0|{\bf \Pi}^2 + {\bf B}_{A}^2|0 \rangle$, Eq. (3.28),
but not renormalized by a counterterm.
We regulate this
value by subtracting the perturbative contribution with $\omega({\bf k})=k$
yielding
\begin{eqnarray}
\langle \frac{\alpha_s}{\pi}F^a_{\mu\nu}F^a_{\mu\nu} \rangle =
\frac{N_c^2-1}{\pi^3}\int_0^{\infty}dk k^2 \alpha_s
\frac{(\omega({\bf k})-k)^2}{2\omega({\bf k})}
\,.\label{eq:3.5}\end{eqnarray}
Using the above dispersion relation, $\omega({\bf k})$,
and a cut-off $\Lambda=4$ $GeV$ that we used in our previous work
\cite{ssjc96},
the computed gluon condensate is  $.013 \, GeV^4 $
in agreement with sum rule results
\cite{ShVaZa}.
The condensate cut-off sensitivity is shown in
Fig. (\ref{fig.4}).  It is interesting that using the same cut-off
as in \cite{ssjc96}, which calculated a condensate of $.012 \, GeV$, produces
a similar result even though the two calculations are quite different.
The logarithmic dependence of the effective gluon mass and condensate can
both be absorbed by the running
coupling constant $\alpha_s(\Lambda)$.

\subsection{Glueball Bound State Equation}

We model the glueball bound state as two
valence constituent gluons.
The glueball wave function in the rest frame is
\begin{eqnarray}
|\psi_n\rangle=\int{d{\bf q}\over(2\pi)^3}\phi_n^{ij}({\bf q})
a_i^{a\dagger}({\bf q})a_j^{a\dagger}(-{\bf q})|0\rangle
\,,\label{eq:3.6}\end{eqnarray}
where repeated indices are summed and $a_i^{a\dagger}({\bf q})$ creates a
quasiparticle from the nontrivial vacuum $|0\rangle$
with energy $\omega({\bf q})$.
Since in this constituent basis our effective Hamiltonian
${ H}_{eff}$ is block-diagonal, mixing with three gluon and
higher states
is suppressed. Therefore we can directly diagonalize in the two gluon $P$
space
and our bound state equation follows by projecting the fundamental equation
${ H}_{eff}|\psi_n\rangle={ E}_n|\psi_n\rangle$
on the $P$ space
\begin{eqnarray}
\langle\psi_n|[H_{eff},a_i^{a\dagger}({\bf q})a_j^{a\dagger}(-{\bf
q})]|0\rangle
=(E_n-E_0)X^{ij}_{i'j'}({\bf q})\phi_n^{i'j'}({\bf q})
\,,\label{eq:3.7}\end{eqnarray}
where
\begin{eqnarray}
X^{ij}_{i'j'}({\bf q})\phi_n^{i'j'}({\bf q})
= D_{ii'}({\bf q})D_{jj'}({\bf q})\phi_n^{i'j'}({\bf q})
+ D_{ij'}({\bf q})D_{ji'}({\bf q})\phi_n^{i'j'}(-{\bf q})
\,.\label{eq:3.8}\end{eqnarray}
The polarization term, $D_{ij}$, is given by Eq. (\ref{eq:2.12})
and $H_{eff}$ by Eq. (\ref{eq:2.21a}),
with matrix elements given in Eq. (\ref{eq:2.22}).
Equation (\ref{eq:3.7}) determines the glueball mass, $M_n = E_n - E_0$, where
$E_0$ is the vacuum energy
defined by $H_{eff}|0\rangle=E_0|0\rangle$.

Using our $H_{eff}$ we have evaluated Eqs. (4.7,4.8) and summarize in
Appendix C the complete
form of the bound state equation with all possible terms through second order.
As mentioned in the introduction and important to repeat,
even though our analysis has the appearance of the TDA, we do not
make this approximation (truncation to the two gluon sector).
Quite the contrary, we include the truncation corrections
which introduces complicated new terms in the bound state equation
as detailed in Appendix C and below.

As a numerical application we consider here only the dominant,
instantaneous part
of the kernel, neglecting perturbative transverse gluon exchange and terms
>from the four-gluon vertex.
The bound state equation for glueball states
having total angular momentum, $J$, parity, $P$, and charge conjugation, $C$,
is then
\begin{eqnarray}
& & M_n\phi_n(q)
= \left[\left(\frac{q^2+\widetilde{m}_{CT}^2(\Lambda)}
{\omega_{\bf q}}+\omega_{\bf q} \right)\right.
\nonumber\\
&+& \left.\frac{1}{4} N_c
\int{d{\bf p}\over(2\pi)^3}
\widetilde{V}_{L+C}({\bf p}-{\bf q})\left(1+({\hat p}{\hat q})^2 \right)
\frac{\omega_{\bf p}^2+\omega_{\bf q}^2}{\omega_{\bf p}\omega_{\bf q}}
{\rm e}^{-p^2/\Lambda^2}
\right]\phi_n(q)
\nonumber\\
&-& \frac{1}{8} N_c \int{d{\bf p}\over(2\pi)^3}
\widetilde{V}_{L+C}({\bf p}-{\bf q})
\frac{(\omega_{\bf p}+\omega_{\bf q})^2}{\omega_{\bf p}\omega_{\bf q}}
F^{JPC}({\bf p},{\bf q})\phi_n(p)
\,.\label{eq:3.9}\end{eqnarray}
The spectroscopic terms, $F^{JPC}$, for the scalar, $0^{++}$, and
pseudoscalar, $0^{-+}$, states are
\begin{eqnarray}
F^{0++}({\bf p},{\bf q}) &=& 1+({\hat p}{\hat q})^2
\,,\nonumber\\
F^{0-+}({\bf p},{\bf q}) &=& 2({\hat p}{\hat q})
\,.\label{eq:3.10}\end{eqnarray}
Here the Coulomb counterterm, $\widetilde{m}_{CT}(\Lambda)$,
is given by Eq. (\ref{eq:3.3}) and $\widetilde{V}_{L+C}$
is the sum of linear and Coulomb potentials in momentum space.

Besides the UV divergences from the Coulomb interaction,
the bound state equation, Eq. (\ref{eq:3.9}), also has
an IR singularity from the confining potential.
The UV divergences in the first integral
are regulated by the exponential cut-off function
and the leading divergent part
is canceled by the mass counterterm $\widetilde{m}_{CT}(\Lambda)$.
The UV behavior of the second integral
is regulated by the wave function, which decreases with increasing momenta.
Although the kinetic (self-energy term) and
potential parts both contribute IR
divergent pieces, the bound state equation is IR finite
due to complete cancelation for small momenta.
This cancelation only occurs for the color singlet state \cite{adler}.

Numerical solutions of Eq. (\ref{eq:3.9})
are obtained variationally with a set of gaussian test functions.
Results for the ground state scalar and pseudoscalar glueball masses,
and first excited states, are
presented in Table \ref{tab.1} and compared with recent quenched lattice
data \cite{lattice}, generated with an improved SII action and
anisotropic lattice.  It is more consistent to compare with quenched lattice
results, rather than improved unquenched treatments, since we also omit the
quark
sector.  Notice agreement is better in the scalar channel and that the mass of
this state is about twice the constituent gluon mass.

\section{Conclusions and Outlook}

In this work we have described in detail a systematic procedure for solving
a bound state problem in QCD. Our study is motivated by the success
of the constituent quark model. Central to our approach are the flow
equations,  which we have applied perturbatively to a QCD motivated Hamiltonian
with confining interaction.  The flow equations generated a block
diagonal, effective
Hamiltonian which provides
an improved framework for vacuum and excited hadron state investigations.
In the block diagonal process, the effective Hamiltonian was also renormalized
through second order in the coupling constant.  In particular, we determined
the appropriate mass counterterm to cancel the
leading UV divergences.

Applications to the gluon vacuum and excited glueball spectrum
produced a nonlinear gap equation and a bound state equation
that is superior to the TDA.
Numerical solutions reproduced the QCD
sum rule gluon condensate value and also provided reasonable agreement
with quenched lattice glueball calculations.  Our results support the
constituent picture for gluonia.

While we have only considered pure gluodynamics in this work,
the same approach can be directly applied
to the full QCD Hamiltonian, including dynamical quarks, where the
issue of chiral symmetry breaking is present.  Of keen interest
will be the structure of the full gap and bound state equations in the combined
quark and gluon sectors.  In particular, application to exotic
four quark and hybrid systems will provide significant new
information, especially the degree of mixing between
$q\bar q$, $qq\bar q \bar q$, $q\bar q g$ and $gg$ states.
Much of this work is in progress and will be reported
in a future publication.

\vspace{1cm}
\acknowledgments
This work was supported by the U.S. DOE
under contracts DE-FGO2-96ER40947 and DE-FGO2-97ER41048. The North Carolina
Supercomputing  Center and the National Energy Research Scientific Computing
Center are also acknowledged for supercomputing time.

\appendix
\section{Complete QCD Motivated Hamiltonian}

Here we specify, in second quantized form, the full QCD motivated Hamiltonian
of gluodynamics in the Coulomb gauge, decomposed in the basis
Eq. (\ref{eq:2.10}). One-body operators and condensate terms arise from normal
ordering with respect to the trial vacuum state $|0\rangle$.
The upper index over a Hamiltonian operator (e.g. $K^{(0)}$, $H^{(1)}$)
denotes the order (power) in (bare) coupling constant.

{\bf Free gluon part:} Eq. (\ref{eq:2.5}) includes the
gluon kinetic energy
\begin{eqnarray}
 K &=& \frac{1}{2}\int\frac{d{\bf k}}{(2\pi)^3}
\left[(\frac{{\bf k}^2}{\omega_{\bf k}}+\omega_{\bf k})
a_i^{a\dagger}({\bf k})a_i^a({\bf k})
+(\frac{{\bf k}^2}{\omega_{\bf k}}-\omega_{\bf k})
\frac{1}{2}( a_i^a({\bf k})a_i^a(-{\bf k})+ {\rm h.c.} ) \right] \nonumber\\
&=& \int\frac{d{\bf k}}{(2\pi)^3}\omega_{\bf k}
a_i^{a\dagger}({\bf k})a_i^a({\bf k})
+\frac{1}{2}\int\frac{d{\bf k}}{(2\pi)^3}
(\frac{{\bf k}^2}{\omega_{\bf k}}-\omega_{\bf k}) \nonumber\\
& \times &
(a_i^{a\dagger}({\bf k})a_i^a({\bf k})
+\frac{1}{2}( a_i^a({\bf k})a_i^a(-{\bf k})+ {\rm h.c.} ))
= K^{(0)} + ( K - K^{(0)})
\,,\label{eq:4.1}\end{eqnarray}
with
\begin{eqnarray}
 K^{(0)} &=& \int\frac{d{\bf k}}{(2\pi)^3}\omega_{\bf k}
a_i^{a\dagger}({\bf k})a_i^a({\bf k})
\,,\label{eq:4.2}\end{eqnarray}
and condensate term
\begin{eqnarray}
 O &=& \frac{1}{2}(N_c^2-1){\bf V}
\int\frac{d{\bf k}}{(2\pi)^3}(\frac{{\bf k}^2}{\omega_{\bf k}}+
\omega_{\bf k})
\nonumber\\
 &=&  (N_c^2-1){\bf V}
\int\frac{d{\bf k}}{(2\pi)^3}\omega_{\bf k} +
\frac{1}{2}(N_c^2-1){\bf V}
\int\frac{d{\bf k}}{(2\pi)^3}(\frac{{\bf k}^2}{\omega_{\bf k}}-
\omega_{\bf k}) =O^{(0)}+  (O - O^{(0)})
\,,\label{eq:4.3}\end{eqnarray}
with
\begin{eqnarray}
 O^{(0)} &=& (N_c^2-1){\bf V}
\int\frac{d{\bf k}}{(2\pi)^3}\omega_{\bf k}
\,,\label{eq:4.3a}\end{eqnarray}
and volume ${\bf V}=(2\pi)^3\delta^{(3)}(0)$.

{\bf Instantaneous interaction:} Eq. (\ref{eq:2.9a}) includes
the linear confining and Coulomb interactions
\begin{eqnarray}
H_{L+C}^{(2)}& & = -{1\over8} f^{abc}f^{ade}
\int\left(\prod_{n=1}^{4} {d{\bf k}_n\over(2\pi)^3}\right)
(2\pi)^3\delta^{(3)}(\sum_m {\bf k}_m)
\left({\omega_2\omega_4\over\omega_1\omega_3}\right)^{1/2}
\widetilde{V}_{L+C}({\bf k}_1+{\bf k}_2)  \label{eq:4.4} \\
& &\hspace{-2cm}
\times {\bf :}
\left[ a^b_i({\bf k}_1) + {a^b_i}^\dagger(-{\bf k}_1)\right]
\left[ a^c_i({\bf k}_2) - {a^c_i}^\dagger(-{\bf k}_2)\right]
\left[ a^d_j({\bf k}_3) + {a^d_j}^\dagger(-{\bf k}_3)\right]
\left[ a^e_j({\bf k}_4) - {a^e_j}^\dagger(-{\bf k}_4)\right]
{\bf :} \nonumber
\,,\end{eqnarray}
where the supercript $(2)$ refers to the Coulomb term, which is second order,
and $\widetilde{V}_{L+C}({\bf k})$ is
\begin{eqnarray}
\widetilde{V}_{L+C}({\bf k}) = 2\pi  C_{adj} {\alpha_s \over k^2}
+ 4\pi {\sigma \over k^4}
\,.\end{eqnarray}
Terms arising from normal ordering are
the one-body operator
\begin{eqnarray}
\Pi_{L+C} = {N_c\over4} & & \int
{d{\bf k}\thinspace d{\bf q} \over(2\pi)^6}
\widetilde{V}_{L+C}({\bf k}+{\bf q})
\left({\omega_{\bf q}^2+\omega_{\bf k}^2 \over
\omega_{\bf k}\omega_{\bf q}}\right)
D_{ij}({\bf q})
\left[ {a^a_i}^\dagger({\bf k}) a^a_j({\bf k})\right] \label{eq:4.5} \\
+ {N_c\over8} & & \int
{d{\bf k}\thinspace d{\bf q} \over(2\pi)^6}
\widetilde{V}_{L+C}({\bf k}+{\bf q})
\left({\omega_{\bf q}^2
- \omega_{\bf k}^2\over\omega_{\bf k}\omega_{\bf q}}\right)
D_{ij}({\bf q})
\left[a^a_i({\bf k}) a^a_j({\bf k}) + {\rm h.c.}\right]  \nonumber
\,,\end{eqnarray}
and the condensate term
\begin{eqnarray}
O_{L+C} &=& \frac{1}{8}N_c(N_c^2-1){\bf V}
\int{d{\bf k}\thinspace d{\bf q}\over(2\pi)^6}
\widetilde{V}_{L+C}({\bf k}+{\bf q})
\left(\frac{\omega_{\bf q}}{\omega_{\bf k}}-1\right)
\left(1+({\hat k}{\hat q})^2\right)
\,.\label{eq:4.6}\end{eqnarray}

{\bf Nonabelian gluon part:} Eq. (\ref{eq:2.9}) includes
to order $O(g)$ the triple-gluon coupling
\begin{eqnarray}
H_{3g}^{(1)} &=& \frac{ig}{2\sqrt{2}}f^{abc}
\int\left(\prod_{n=1}^{3} {d{\bf k}_n\over(2\pi)^3}\right)
(2\pi)^3\delta^{(3)}(\sum_m {\bf k}_m)
\frac{k_{1j}}{\sqrt{\omega_1\omega_2\omega_3}} \label{eq:4.7} \\
& & \times {\bf :}
\left[a^a_i({\bf k}_1) + {a^a_i}^\dagger(-{\bf k}_1)\right]
\left[a^b_j({\bf k}_2) + {a^b_j}^\dagger(-{\bf k}_2)\right]
\left[a^c_i({\bf k}_3) + {a^c_i}^\dagger(-{\bf k}_3)\right]
{\bf :}  \nonumber
\,,\end{eqnarray}
where $\omega_1\equiv \omega_{{\bf k}_1}$, etc.
In order $O(g^2)$, the normal ordered four-gluon vertex is
\begin{eqnarray}
H_{4g}^{(2)} & &  ={\alpha_s\pi\over4} f^{abc}f^{ade}
\int\left(\prod_{n=1}^{4} {d{\bf k}_n\over(2\pi)^3}\right)
(2\pi)^3\delta^{(3)}(\sum_m {\bf k}_m)
{1\over\sqrt{\omega_1\omega_2\omega_3\omega_4}} \label{eq:4.8} \\
& &\hspace{-2cm}
\times {\bf :}
\left[ a^b_i({\bf k}_1) + {a^b_i}^\dagger(-{\bf k}_1)\right]
\left[ a^c_j({\bf k}_2) + {a^c_j}^\dagger(-{\bf k}_2)\right]
\left[ a^d_i({\bf k}_3) + {a^d_i}^\dagger(-{\bf k}_3)\right]
\left[ a^e_j({\bf k}_4) + {a^e_j}^\dagger(-{\bf k}_4)\right]
{\bf :} \nonumber
\,.\end{eqnarray}
Also, after normal ordering one obtains a one-body operator
\begin{eqnarray}
\Pi_{4g} &=& \alpha_s\pi N_c
\int{d{\bf k}\thinspace d{\bf q}\over(2\pi)^6}
{1\over\omega_{\bf k}\omega_{\bf q}}[2\delta_{ij} - D_{ij}({\bf q})]
\label{eq:4.9} \\
& &\quad\qquad\times\left[ {a^a_i}^\dagger({\bf k}) a^a_j({\bf k})
+\frac{1}{2}\left(a^a_i({\bf k}) a^a_j(-{\bf k})
+{\rm h.c.}\right)\right]  \nonumber
\,,\end{eqnarray}
and a condensate term
\begin{eqnarray}
O_{4g} &=& \frac{\alpha_s\pi}{4} N_c(N_c^2-1){\bf V}
\int{d{\bf k}\thinspace d{\bf q}\over(2\pi)^6}
{1\over\omega_{\bf k}\omega_{\bf q}}
\left(3-({\hat k}{\hat q})^2 \right)
\,.\label{eq:4.10}\end{eqnarray}
One-body and condensate terms diverge in the UV region
and must be regulated.

\section{Second Order Flow Equations}

Solving the second order flow equation for the particle number
conserving part $H_d$,  Eq. (\ref{eq:2.20}),
generates three types of Hamiltonian operators:
two-body effective interactions, one-body polarization terms and condensates.
We consider each separately.

\subsection{Effective Interaction (two-body sector)}

We calculate an effective gluon interaction in the color singlet channel
for glueball Fock states.
The general form of an effective interaction is given by Eq. (\ref{eq:2.27})
and includes interactions from the original Hamiltonian and
a new interaction, generated to second order by flow equations
in the two-body sector.
Consider first the generated terms.
The $t$-channel diagrams, arising from the commutator
$[a_{1i}^aa_{2j}^{b\dagger}a_{3k}^{c\dagger},a_{1'i'}^{a'\dagger}a_{2'j'}^{b'}a_
{3'k'}^{c'}]$
(diagram without backward motion) and from the commutator
$[a_{1i}^aa_{2j}^{b}a_{3k}^{c},a_{1'i'}^{a'\dagger}a_{2'j'}^{b'\dagger}a_{3'k'}^
{c'\dagger}]$
(Z-graph in t-channel), do not contribute to the color singlet state;
only the $s$-channel diagram, coming from the commutator
$[a_{1i}^{a\dagger}a_{2j}^{b}a_{3k}^{c},a_{1'i'}^{a'}a_{2'j'}^{b'\dagger}a_{3'k'
}^{c'\dagger}]$,
needs to be calculated (here the notation is $a_{1i}^{a}=a_{i}^{a}({\bf k}_1),
etc.)$. In the c.m. frame the flow equation for the effective interaction is
\begin{eqnarray}
\frac{d V_{\rm gen}(l)}{dl}   &=& \frac{1}{8}f^{abc}f^{a'b'c}
\int {d{\bf k} d{\bf q} \over(2\pi)^6}
\frac{1}{\omega_{\bf k}\omega_{\bf q}\omega_{{\bf k}-{\bf q}}}
\nonumber\\
& \times &
\Gamma_{ijk}({\bf k},-{\bf q},-({\bf k}-{\bf q}) )
\Gamma_{i'j'k'}({\bf k},-{\bf q},-({\bf k}-{\bf q}))
D_{kk'}({\bf k}-{\bf q})\cdot 4\cdot 9   \nonumber \\
& \times &  \left[
\eta_1({\bf q},{\bf k},{\bf k}-{\bf q};l)g_1({\bf k},{\bf q},{\bf k}-{\bf
q};l) +
\eta_1({\bf k},{\bf q},{\bf k}-{\bf q};l)g_1({\bf q},{\bf k},{\bf k}-{\bf q};l)
\right]
\nonumber\\
& \times & {\bf :}
{a^b_j}^\dagger({\bf q}){a^{b'}_{j'}}^\dagger(-{\bf q})a^a_i({\bf
k})a^{a'}_{i'}(-{\bf k})
{\bf :}
\,,\label{eq:5.1}\end{eqnarray}
where the factor $4$ is the number of permutations and the property of
$\Gamma$-factors, Eq. (\ref{eq:2.16c}), is used.
The generators are given in Eq. (\ref{eq:2.18}) and coupling constants
in Eq. (\ref{eq:2.19}).
The $\eta g$ sum
corresponds to two different time-ordered s-channel diagrams.
We introduce the factor $S_{ijk,i'j'k'}({\bf k},{\bf q})$ by
\begin{eqnarray}
\Gamma_{ijk}({\bf k},-{\bf q},-({\bf k}-{\bf q}))
\Gamma_{i'j'k'}({\bf k},-{\bf q},-({\bf k}-{\bf q})) =
\frac{1}{6}\frac{1}{6} 4 S_{ijk,i'j'k'}({\bf k},{\bf q})
\,.\label{eq:5.2}\end{eqnarray}
which, due to transversality, can be expressed as
\begin{eqnarray}
S_{ijk,i'j'k'}({\bf k},{\bf q})=(k_j\delta_{ik}-
\frac{1}{2}(k+q)_k\delta_{ij}+q_i\delta_{jk})
(k_{j'}\delta_{i'k'}-
\frac{1}{2}(k+q)_{k'}\delta_{i'j'}+q_{i'}\delta_{j'k'})
\,,\label{eq:5.3}\end{eqnarray}
and is symmetric under interchange of ${\bf k}$ and ${\bf q}$.
The tensor structure of the generated interaction is defined by the
contraction
$S_{ijk,i'j'k'}({\bf k},{\bf q})D_{kk'}({\bf k}-{\bf q})\equiv
T_{ij,i'j'}({\bf k},{\bf q})$,
where
\begin{eqnarray}
T_{ij,i'j'}({\bf k},{\bf q}) &=& \left(\phantom{\frac{1}{2}}\hspace{-0.3cm}
q_iq_{i'}\delta_{jj'}+k_jk_{j'}\delta_{ii'}
+q_i(k_{j'}\delta_{i'j}-k_j\delta_{i'j'})
+q_{i'}(k_{j}\delta_{ij'}-k_{j'}\delta_{ij }) \right.\nonumber\\
&+& \left.\frac{k^2q^2}{({\bf k}-{\bf q})^2}
(1-({\hat k}{\hat q})^2)\delta_{ij}\delta_{i'j'}\right)
\nonumber\\
&=& ({\bf k}-{\bf q})^2
\left( \phantom{\frac{1}{2}}\hspace{-0.3cm}
n_in_{i'}\delta_{jj'}+n_jn_{j'}\delta_{ii'}
+n_in_j\delta_{i'j'}+n_{i'}n_{j'}\delta_{ij}
-n_in_{j'}\delta_{i'j}-n_{i'}n_j\delta_{ij'} \right.
\nonumber\\
&+& \left.\frac{k^2q^2}{({\bf k}-{\bf q})^4}(1-({\hat k}{\hat q})^2)
\delta_{ij}\delta_{i'j'}  \right)
\,,\label{eq:5.4}\end{eqnarray}
and $n_i$ are the components of the unit vector
${\bf n} =({\bf k}-{\bf q})/|{\bf k}-{\bf q}|$. Integrating the
flow equation, Eq. (\ref{eq:5.1}), we obtain the generated interaction
$V_{\rm gen}(l\rightarrow\infty)\equiv V_{\rm gen}$
(the initial value is $V_{\rm gen}(l=0)=0$)
\begin{eqnarray}
V_{\rm gen} &=& \alpha_s 2\pi f^{abc}f^{a'b'c}
\int {d{\bf k} d{\bf q} \over(2\pi)^6}
\frac{1}{\omega_{\bf k}\omega_{\bf q}\omega_{{\bf k}-{\bf q}} }
T_{ij,i'j'}({\bf k},{\bf q})
\frac{D_1+D_{1'}}{D_1^2+D_{1'}^2}
\nonumber\\
& \times &
 {\bf :}
 {a^b_j}^\dagger({\bf q}){a^{b'}_{j'}}^\dagger(-{\bf q})a^a_i({\bf
k})a^{a'}_{i'}(-{\bf k})
{\bf :}
\,,\label{eq:5.5}\end{eqnarray}
with
\begin{eqnarray}
 D_{1}=-(-\omega_{\bf k}+\omega_{\bf q}+\omega_{{\bf k}-{\bf q}}) \,, \,
 D_{1'}=-(-\omega_{\bf q}+\omega_{\bf k}+\omega_{{\bf k}-{\bf q}})
\,.\label{eq:5.6}\end{eqnarray}
This reduces to
\begin{eqnarray}
V_{\rm gen} &=& -\alpha_s 2\pi f^{abc}f^{a'b'c}
\int {d{\bf k} d{\bf q} \over(2\pi)^6}
\frac{1}{\omega_{\bf k}\omega_{\bf q}\omega_{{\bf k}-{\bf q}}^2 }
T_{ij,i'j'}({\bf k},{\bf q})
\left(1-\frac{(\omega_{\bf k}-\omega_{\bf q})^2}{(\omega_{\bf
k}-\omega_{\bf q})^2
+\omega_{{\bf k}-{\bf q}}^2} \right)
\nonumber\\
& \times &
 {\bf :}
 {a^b_j}^\dagger({\bf q}){a^{b'}_{j'}}^\dagger(-{\bf q})a^a_i({\bf
k})a^{a'}_{i'}(-{\bf k})
{\bf :}
\,.\label{eq:5.7}\end{eqnarray}

Now consider the second order gluon interactions
that conserve particle number. The generator
$\eta^{(1)}$, Eq. (\ref{eq:2.17}), produces second order
interactions that change the number of quasiparticles,
for example, an interaction $W$ having form
$a^{a\dagger}_{1i}a^{b}_{2j}a^{c}_{3k}a^{d}_{4l}$.
However, the generator choice, $\eta=\eta^{(1)}+\eta^{(2)}$, eliminates $W$,
where the second order generator is calculated from
$\eta^{(2)}=[K^{(0)},W]$.

To maintain gauge invariance we also include the normal ordered instantaneous,
Eq. (\ref{eq:4.4}), and the four-gluon, Eq. (\ref{eq:4.8}), interactions of the
original Hamiltonian in the color-singlet channel.  Projecting these terms on
the color singlet state yields
\begin{eqnarray}
V_{L+C} &=& -\frac{1}{8}f^{abc}f^{a'b'c}
\int {d{\bf k} d{\bf q} \over(2\pi)^6}
\frac{(\omega_{\bf k}+\omega_{\bf q})^2}{\omega_{\bf k}\omega_{\bf q}}
\widetilde{V}_{L+C}({\bf k}-{\bf q}) \thinspace
\delta_{ij}\delta_{i'j'}
\thinspace {\bf :}
 {a^b_j}^\dagger({\bf q}){a^{b'}_{j'}}^\dagger(-{\bf q})a^a_i({\bf
k})a^{a'}_{i'}(-{\bf k})
{\bf :} \nonumber\\
V_{4g} &=& \frac{\alpha_s\pi}{4}
f^{abc}f^{a'b'c}
\int {d{\bf k} d{\bf q} \over(2\pi)^6}
\frac{2}{\omega_{\bf k}\omega_{\bf q}}
(\delta_{ii'}\delta_{jj'}-\delta_{ij'}\delta_{i'j})
\thinspace {\bf :}
 {a^b_j}^\dagger({\bf q}){a^{b'}_{j'}}^\dagger(-{\bf q})a^a_i({\bf
k})a^{a'}_{i'}(-{\bf k})
{\bf :}
\,,\label{eq:5.8}\end{eqnarray}
The complete effective gluon interaction is given by
\begin{eqnarray}
V_{eff} &=& V_{L+C}+V_{4g}+V_{\rm gen}
\,,\label{eq:5.9}\end{eqnarray}
with the generated interaction defined by Eq. (\ref{eq:5.7}).

\subsection{Polarization Operator (one-body sector)}

Two contraction terms from the commutator $[\eta^{(1)},H_{3g}^{(1)}]$
contribute to one-body operators
\begin{eqnarray}
\Pi=\int\frac{d{\bf k}}{(2\pi)^3}\left( \Pi_{ij}^{ab}({\bf k})
a_{i}^{a\dagger}({\bf k})a_{j}^{b}({\bf k})
+M_{ij}^{ab}({\bf k})\thinspace
\frac{1}{2}\left( a_{i}^{a}({\bf k})a_{j}^{b}(-{\bf k})
           + {\rm h.c.} \right) \right)
\,,\label{eq:5.10}\end{eqnarray}
with polarization amplitudes, $\Pi_{ij}^{ab}({\bf k})$,
$M_{ij}^{ab}({\bf k})$, given by
gluon loop integrals of the form
\begin{eqnarray}
\Pi_{ij}^{ab}({\bf k})=\int\frac{d{\bf q}}{(2\pi)^3}
\widetilde{\Pi}_{ij}^{ab}({\bf k},{\bf q})
\,.\label{eq:5.11}\end{eqnarray}
The specific integrand structure is given below.
Note, from rotational invariance, the one-body operators
$\Pi_{ij}^{ab}({\bf k})$ and $M_{ij}^{ab}({\bf k})$ have tensor structure,
$A^{ab}(k)\delta_{ij}+B^{ab}(k){\hat k}_i{\hat k}_j$, however,
due to transversality, only the term proportional to $\delta_{ij}$
survives after integrating Eq. (\ref{eq:5.10}).
Hence, we only need to extract $A^{ab}(k)$ which can be achieved by
choosing
$k=k_z$ and considering components
$i=j=x$ and $y$ but $\neq$ $z$. Then in Eq. (\ref{eq:5.11}), where
$\widetilde{\Pi}_{ij}^{ab}({\bf k},{\bf q})$ also has terms of the type
$\delta_{ij}$ and ${\hat q}_i{\hat q}_j$ (there are no terms proportional to
${\hat k}_i$ and
${\hat k}_j$), one can substitute
\begin{eqnarray}
{\hat q}_i{\hat q}_j\rightarrow\frac{1}{2}(1-({\hat k}{\hat q})^2)\delta_{ij}
\,,\label{eq:5.12}\end{eqnarray}
to extract
the $\delta_{ij}$ component in Eq. (\ref{eq:5.11}).
>From the second order flow equation, Eq. (\ref{eq:2.20}),
the gluon polarization operator is
\begin{eqnarray}
& & \frac{d\Pi(l)}{dl} = \frac{1}{8}N_c\delta^{aa'}
\int {d{\bf k} d{\bf q} \over(2\pi)^6}
\frac{1}{\omega_{\bf k}\omega_{\bf q}\omega_{{\bf k}-{\bf q}}}
\nonumber\\
&\times&
\Gamma_{ijk}({\bf k},-{\bf q},-({\bf k}-{\bf q}))
\Gamma_{i'j'k'}({\bf k},-{\bf q},-({\bf k}-{\bf q}))
D_{jj'}({\bf q})D_{kk'}({\bf k}-{\bf q})\cdot 2\cdot 9\cdot 2
\nonumber\\
& & \left[
\left(\eta_1({\bf k},{\bf q},{\bf k}-{\bf q};l)g_1({\bf k},{\bf q},{\bf
k}-{\bf q};l)
+\eta_0({\bf k},{\bf q},{\bf k}-{\bf q};l)g_0({\bf k},{\bf q},{\bf k}-{\bf
q};l)\right)
\thinspace
a^{a\dagger}_i({\bf k})a^{a'}_{i'}({\bf k})
\right. \nonumber \\
&+& \left.
\left(\eta_1({\bf k},{\bf q},{\bf k}-{\bf q};l)g_0({\bf k},{\bf q},{\bf
k}-{\bf q};l)
+\eta_0({\bf k},{\bf q},{\bf k}-{\bf q};l)g_1({\bf k},{\bf q},{\bf k}-{\bf
q};l)\right)\right.
\label{eq:5.13}\\
&\times & \left. \frac{1}{2}(a^{a}_i({\bf k})a^{a'}_{i'}(-{\bf k})+{\rm h.c.})
\right] \nonumber
\,,\end{eqnarray}
where $\eta_1g_1$ corresponds to a gluon loop without backward motion,
$\eta_0g_0$, a $Z$-graph, and the last two terms $\eta_1g_0$ and
$\eta_0g_1$, are gluon loop diagrams with two incoming and two outgoing gluon
lines. In calculating Eq. (\ref{eq:5.13}) we have also used the property,
Eq. (\ref{eq:2.16c}),  for $\Gamma$-factors.

The polarization operator tensor structure is given by a double contraction
of the $S({\bf k},{\bf q})$ factor, Eq. (\ref{eq:5.3}), with polarization
terms
in Eq. (\ref{eq:5.13}). We then define
\begin{eqnarray}
F_{ii'}({\bf k},{\bf q}) \equiv
S_{ijk,i'j'k'}({\bf k},{\bf q})D_{jj'}({\bf q})D_{kk'}({\bf k}-{\bf q})
\nonumber\\
=\delta_{ii'}k^2q^2\left( (\frac{1}{q^2}+\frac{1}{({\bf k}-{\bf q})^2})
(1-({\hat k}{\hat q})^2)  \right)
&+& 2q_iq_{i'}\left( 1-\frac{k^2}{2({\bf k}-{\bf q})^2}(1-({\hat k}{\hat
q})^2)
\right)
\,,\label{eq:5.14}\end{eqnarray}
with repeated indices summed.
Due to transversality, only the $\delta_{ii'}$ component
is retained, therefore from Eq. (\ref{eq:5.12})
\begin{eqnarray}
& & F_{ii'}({\bf k},{\bf q})\rightarrow \delta_{ii'}G({\bf k},{\bf q})/2 \,,
\nonumber\\
& & G({\bf k},{\bf q})=2(1-({\hat k}{\hat q})^2)
\left( k^2+q^2+\frac{k^2q^2}{2({\bf k}-{\bf q})^2}
(1+({\hat k}{\hat q})^2) \right)
\,.\label{eq:5.15}\end{eqnarray}
We integrate the flow equation,
Eq. (\ref{eq:5.13}), with the generators and coupling constants given by
Eq. (\ref{eq:2.18})  and Eq. (\ref{eq:2.19}), respectively,
producing the second order polarization operator correction,
$\delta\Pi = \Pi(l)-\Pi(l_0=0)$. The polarization operator, at a scale
$l=1/\lambda^2$, is then
\begin{eqnarray}
\Pi_{\rm gen}(\lambda) &=& \alpha_s\pi N_c\int {d{\bf k} d{\bf q}
\over(2\pi)^6}
\frac{1}{\omega_{\bf k}\omega_{\bf q}\omega_{{\bf k}-{\bf q}} }
G({\bf k},{\bf q})\nonumber\\
&\times& \left[
\left(\frac{1}{2D_0}{\rm e}^{-2D_0^2/\lambda^2}
+\frac{1}{2D_1}{\rm e}^{-2D_1^2/\lambda^2} \right)
a^{a\dagger}_i({\bf k})a^{a}_{i}({\bf k}) \right.
\nonumber\\
&+& \left. \frac{D_0+D_1}{D_0^2+D_1^2}
{\rm e}^{-(D_0^2+D_1^2)/\lambda^2}
\frac{1}{2}(a^{a}_i({\bf k})a^{a}_{i}({-\bf k})+{\rm h.c.})
\right]
\,,\label{eq:5.16}\end{eqnarray}
with
\begin{eqnarray}
 D_0=-(\omega_{\bf k}+\omega_{\bf q}+\omega_{{\bf k}-{\bf q}}) \,, \,
 D_1=-(-\omega_{\bf k}+\omega_{\bf q}+\omega_{{\bf k}-{\bf q}})
\,.\label{eq:5.17}\end{eqnarray}
For large momenta flowing in the loop, one has
$D_0\sim D_1\sim -(\omega_{\bf q}+\omega_{{\bf k}-{\bf q}})\sim
-2\omega_{\bf q}$.
Finally, rescaling the cut-off,
$\lambda \rightarrow\sqrt{2}\lambda$, which does not change results,
the polarization operator generated by the flow equations through second
order is
\begin{eqnarray}
\Pi_{\rm gen}(\lambda) &=& -\alpha_s\pi N_c
\int {d{\bf k} d{\bf q} \over(2\pi)^6}
\frac{G({\bf k},{\bf q})}{\omega_{\bf k}\omega_{\bf q}\omega_{{\bf k}-{\bf
q}}}
\frac{1}{\omega_{\bf q}+\omega_{{\bf k}-{\bf q}} }
{\rm e}^{-4q^2/\lambda^2}
\nonumber\\
& \times &
\left( a^{a\dagger}_i({\bf k})a^{a}_{i}({\bf k})
+\frac{1}{2}(a^{a}_i({\bf k})a^{a}_{i}({-\bf k})+{\rm h.c.})
\right)
\,,\label{eq:5.18}\end{eqnarray}
where in the exponential factor we have used the
free dispersion relation, $\omega_{\bf q}=q$, valid for large cut-off values.
We then calculate one-body operators, by normal ordering the instantaneous
interactions and the four-gluon vertex, Eq. (\ref{eq:4.4}) and Eq.
(\ref{eq:4.8}),
respectively, and again extract the $\delta_{ij}$ component
using Eq. (\ref{eq:5.12}).
The regulated polarization operators are
\begin{eqnarray}
\Pi_{L+C}(\lambda) &=& \frac{1}{8}N_c\int {d{\bf k} d{\bf q} \over(2\pi)^6}
\frac{1}{\omega_{\bf k}\omega_{\bf q}}
\widetilde{V}_{L+C}({\bf k}-{\bf q})(1+({\hat k}{\hat q})^2){\rm
e}^{-q^2/\lambda^2}
\nonumber\\
&\times & \left( (\omega_{\bf q}^2+\omega_{\bf k}^2)
a^{a\dagger}_i({\bf k})a^{a}_{i}({\bf k})
+(\omega_{\bf q}^2-\omega_{\bf k}^2)
\frac{1}{2}(a^{a}_i({\bf k})a^{a}_{i}({-\bf k})+{\rm h.c.}) \right)
\label{eq:5.19} \\
\Pi_{4g}(\lambda) &=& \frac{\alpha_s\pi}{2}N_c\int {d{\bf k} d{\bf q}
\over(2\pi)^6}
\frac{1}{\omega_{\bf k}\omega_{\bf q}}(3-({\hat k}{\hat q})^2)
{\rm e}^{-q^2/\lambda^2}
\nonumber\\
&\times & \left(a^{a\dagger}_i({\bf k})a^{a}_{i}({\bf k})
+\frac{1}{2}(a^{a}_i({\bf k})a^{a}_{i}({-\bf k})+{\rm h.c.}) \right) \nonumber
\,.\end{eqnarray}
We choose the same regulating function
to match energy denominators.
Note that in the generated term the regulator emerges naturally
>from the flow equations, while the normal ordered
terms require introducing a loop integral regulator with form consistent
with the generated term.
The complete polarization operator is then
\begin{eqnarray}
\Pi(\lambda)=\Pi_{L+C}(\lambda)+\Pi_{4g}(\lambda)+\Pi_{\rm gen}(\lambda)
\,.\label{eq:5.20}\end{eqnarray}

We conclude this subsection by specifying the counterterm
for the leading UV divergences in the polarization operator.
The instantaneous (Coulomb only), four-gluon and generated terms
all contribute to the quadratically UV divergent part, $\Pi^{div}$, with
respective weights, $1, 2, -1/4$
\begin{eqnarray}
\Pi^{div}(\Lambda) &=&
\frac{\alpha_s}{\pi}\frac{2N_c}{3}\Lambda^2\left(1+2-\frac{1}{4}\right)
\nonumber \\
&\times&
\int {d{\bf k} \over(2\pi)^3} \frac{1}{2\omega_{\bf k}}
\left(a^{a\dagger}_i({\bf k})a^{a}_{i}({\bf k})
+\frac{1}{2}(a^{a}_i({\bf k})a^{a}_{i}({-\bf k})+{\rm h.c.}) \right)
\,.\label{eq:5.21}\end{eqnarray}
The appropriate counterterm is then
\begin{eqnarray}
\delta X^{\prime}_{CT}(\Lambda) &=&  \int {d{\bf k} \over(2\pi)^3}
\frac{m_{CT}^2(\Lambda)}{2\omega_{\bf k}}
\left(a^{a\dagger}_i({\bf k})a^{a}_{i}({\bf k})
+\frac{1}{2}(a^{a}_i({\bf k})a^{a}_{i}({-\bf k})+{\rm h.c.}) \right)
\nonumber\\
&=& \frac{m_{CT}^2(\Lambda)}{2}\int d{\bf x}
{\bf :}  A_i^a({\bf x})A_i^a({\bf x}) {\bf :}
\,,\label{eq:5.22}\end{eqnarray}
with mass
\begin{eqnarray}
m_{CT}^2(\Lambda)= -\frac{\alpha_s}{\pi}N_c\frac{11}{6}\Lambda^2
\,.\label{eq:5.23}\end{eqnarray}
When the quark sector is included using this same procedure,
the counterterm coefficient reproduces
the QCD $\beta$-function. This result was also obtained
in our previous study \cite{robertson}.

\subsection{Gluon Condensate (zero-body sector)}

The second order flow equation, Eq. (\ref{eq:2.20}), for the condensate term is
\begin{eqnarray}
& & \frac{dO_{\rm gen}(l)}{dl} = \frac{1}{8}N_c(N_c^2-1){\bf V}
\int {d{\bf k} d{\bf q} \over(2\pi)^6}
\frac{1}{\omega_{\bf k}\omega_{\bf q}\omega_{{\bf k}-{\bf q}}}
\nonumber\\
& & \times
\Gamma_{ijk}({\bf k},-{\bf q},-({\bf k}-{\bf q}))
\Gamma_{i'j'k'}({\bf k},-{\bf q},-({\bf k}-{\bf q}))
D_{ii'}({\bf k})D_{jj'}({\bf q})D_{kk'}({\bf k}-{\bf q})\cdot 6 \cdot 2
\nonumber\\
& & \times
\eta_0({\bf k},{\bf q},{\bf k}-{\bf q};l)g_0({\bf k},{\bf q},{\bf k}-{\bf q};l)
\,,\label{eq:5.24}\end{eqnarray}
where only the commutator
$[a_{1i}^aa_{2j}^{b}a_{3k}^{c},a_{1'i'}^{a'\dagger}a_{2'j'}^{b'\dagger}a_{3'k'}^
{c'\dagger}]$
contributes to the vacuum expectation value. The factor of $6$ is the number of
permutations and the volume is ${\bf V}=(2\pi)^3\delta^{(3)}(0)$.

The tensor structure of the condensate term is given by the triple contraction
of the $S$-factor, Eq. (\ref{eq:5.3}), with polarization terms, i.e.
$S_{ijk,i'j'k'}({\bf k},{\bf q})D_{ii'}({\bf k})D_{jj'}({\bf
q})D_{kk'}({\bf k}-{\bf q})\equiv
G({\bf k},{\bf q})$, where the function $G({\bf k},{\bf q})$ is given by Eq.
(\ref{eq:5.15}). Integrating yields the
correction, $\delta O=O(l)-O(l_0=0)$, where $O(l)$ is a condensate
for flow parameter $l$, related to the energy scale $\lambda$ by
$l=1/\lambda^2$.  The resulting
generated condensate term through second order is
\begin{eqnarray}
O_{\rm gen}(\lambda) = \alpha_s\pi N_c(N_c^2-1){\bf V}
\int {d{\bf k} d{\bf q} \over(2\pi)^6}
\frac{1}{\omega_{\bf k}\omega_{\bf q}\omega_{{\bf k}-{\bf q}} }
\frac{G({\bf k},{\bf q})}{3 D_0}
{\rm e}^{-2D_0^2/\lambda^2}
\,,\label{eq:5.25}\end{eqnarray}
where the energy difference $D_0$ is given by Eq. (\ref{eq:5.17}).
\footnote{
In the gap equation calculations we use the symmetrized form
\begin{eqnarray}
O_{\rm gen}(\lambda) &=& \alpha_s\pi N_c(N_c^2-1){\bf V}
\int {d{\bf k}_1 d{\bf k}_2 d{\bf k}_3 \over(2\pi)^9}
\frac{1}{\omega_1\omega_2\omega_3 }
\nonumber\\
&\times & (2\pi)^3\delta^{(3)}({\bf k}_1+{\bf k}_2+{\bf k}_3)
\frac{\widetilde{G}({\bf k}_1,{\bf k}_2,{\bf k}_3)}{3 D_0}
{\rm e}^{-2D_0^2/\lambda^2} \nonumber
\,,\label{eq:5.26}\end{eqnarray}
where $D_0$ is given in Eq. (\ref{eq:2.18a}), and
\begin{eqnarray}
& & \widetilde{G}({\bf k}_1,{\bf k}_2,{\bf k}_3) =
\left( (k_1^2+k_2^2+k_3^2)
(1-2 ({\hat k}_1{\hat k}_2) ({\hat k}_1{\hat k}_3) ({\hat k}_2{\hat k}_3)
)\right.
\nonumber\\
& & \left. +k_1^2 ({\hat k}_1{\hat k}_2)^2({\hat k}_1{\hat k}_3)^2
    +k_2^2 ({\hat k}_1{\hat k}_2)^2({\hat k}_2{\hat k}_3)^2
+k_3^2 ({\hat k}_1{\hat k}_3)^2({\hat k}_2{\hat k}_3)^2
 \right)
\nonumber\\
& & \times (2\pi)^3\delta^{(3)}({\bf k}_1+{\bf k}_2+{\bf k}_3)
 \rightarrow G({\bf k},{\bf q}) \nonumber
\,.\label{eq:5.27}\end{eqnarray}
}
The three regulated condensate terms, instantaneous,
Eq. (\ref{eq:4.6}), four-gluon, Eq. (\ref{eq:4.10}), and
generated can be summarized
\begin{eqnarray}
O_{L+C}(\lambda) &=& \frac{1}{8} N_c(N_c^2-1){\bf V}
\int{d{\bf k}\thinspace d{\bf q}\over(2\pi)^6}
\widetilde{V}({\bf k}-{\bf q})
\left(\frac{\omega_{\bf q}}{\omega_{\bf k}}-1\right)
\left(1+({\hat k}{\hat q})^2\right){\rm e}^{-(|q|+|k|)^2/\lambda^2}
\nonumber\\
O_{4g}(\lambda) &=& \frac{\alpha_s\pi}{4} N_c(N_c^2-1){\bf V}
\int{d{\bf k}\thinspace d{\bf q}\over(2\pi)^6}
{1\over\omega_{\bf k}\omega_{\bf q}}
\left(3-({\hat k}{\hat q})^2 \right){\rm e}^{-(|q|+|k|)^2/\lambda^2}
\label{eq:5.28} \\
O_{\rm gen}(\lambda) &=& -\alpha_s\pi N_c(N_c^2-1){\bf V}
\int {d{\bf k} d{\bf q} \over(2\pi)^6}
\frac{1}{3\omega_{\bf k}\omega_{\bf q}\omega_{{\bf k}-{\bf q}} }
\frac{G({\bf k},{\bf q})}{(\omega_{\bf k}+\omega_{\bf q}+\omega_{{\bf
k}-{\bf q}}) }
{\rm e}^{-(|q|+|k|+|k-q|)^2/\lambda^2} \nonumber
\,.\end{eqnarray}
The regulating procedure is the same as above.
The complete radiative correction to the gluon condensate,
Eq. (\ref{eq:4.3}), is then
\begin{eqnarray}
O(\lambda)=O_{L+C}(\lambda)+O_{4g}(\lambda)+O_{\rm gen}(\lambda)
\,.\label{eq:5.29}\end{eqnarray}

Finally, we determine the counterterm for the leading UV
divergences in the zero-body sector. Again, the instantaneous
(only Coulomb part), four-gluon and
generated terms contribute, respectively
\begin{eqnarray}
O^{\rm div} (\Lambda) &=&
\frac{\alpha_s}{\pi}\frac{2N_c(N_c^2-1){\bf V}}{3}\Lambda^2
\left(1+2-\frac{1}{4}\right)
\int {d{\bf k} \over(2\pi)^3} \frac{1}{2\omega_{\bf k}}
\,.\label{eq:5.30}\end{eqnarray}
Hence the counterterm is
\begin{eqnarray}
\delta X^{\prime\prime}_{CT}(\Lambda) &=&  (N_c^2-1){\bf V}
\int {d{\bf k} \over(2\pi)^3}
\frac{m_{CT}^2(\Lambda)}{2\omega_{\bf k}}
= \frac{m_{CT}^2(\Lambda)}{2}\int d{\bf x}
\langle 0| A_i^a({\bf x})A_i^a({\bf x})|0\rangle
\,,\label{eq:5.31}\end{eqnarray}
where the mass $m_{CT}(\Lambda)$ is given by Eq. (\ref{eq:5.23}).
Combining the counterterms from one-body, Eq. (\ref{eq:5.22}), and
zero-body, Eq. (\ref{eq:5.31}), sectors yields
\begin{eqnarray}
\delta X_{CT} &=& \delta X^{\prime}_{CT}+\delta X^{\prime\prime}_{CT}
= m_{CT}^2(\Lambda) {\rm Tr}\int d{\bf x}{\bf A}^2({\bf x})
\,,\label{eq:5.32}\end{eqnarray}
which is the complete mass counterterm in the effective Hamiltonian,
renormalized
through second order, Eq. (\ref{eq:2.21a}).
Note, that the one- and zero-body counterterms have the same mass coefficient,
$m_{CT}(\Lambda)$. This suggests one can work directly
in a field theoretical basis without a Fock operator decomposition.

\section{Complete Glueball Bound State Equation}

In this appendix we present the complete bound state equation,
including all second order terms.
Projecting the effective Hamiltonian,
Eq. (\ref{eq:2.21a}), on the two-gluon Fock basis, Eq. (\ref{eq:3.7}),
generates the bound state equation for a glueball at rest with
mass $M = E - E_0$ (we suppress the excited state quantum index number)
\begin{eqnarray}
& & M \left( X^{ij}_{ln}({\bf q})\phi_{+}^{ln}({\bf q})
+Y^{ij}_{ln}({\bf q})\phi_{-}^{ln}({\bf q}) \right)
= \left[\left(\frac{q^2+m_{CT}^2(\Lambda)}{\omega_{\bf q}}+\omega_{\bf q}
\right)\right.
\nonumber\\
&+& \left.\frac{1}{4} N_c
\int{d{\bf p}\over(2\pi)^3}
\widetilde{V}_{L+C}({\bf p}-{\bf q})\left(1+({\hat p}{\hat q})^2 \right)
\frac{\omega_{\bf p}^2+\omega_{\bf q}^2}{\omega_{\bf p}\omega_{\bf q}}
{\rm e}^{-p^2/\Lambda^2} \right.
\nonumber\\
&+&\left. \alpha_s\pi N_c
\int{d{\bf p}\over(2\pi)^3}{1\over\omega_{\bf p}\omega_{\bf q}}
\left(3-({\hat k}{\hat q})^2 \right)
{\rm e}^{-p^2/\Lambda^2}    \right.
\nonumber\\
&-& \left. \alpha_s\pi N_c
\int{d{\bf p}\over(2\pi)^3}{1\over\omega_{\bf p}\omega_{\bf q}\omega_{{\bf
p}-{\bf q}} }
\frac{G({\bf p},{\bf q})}{\omega_{\bf p}+\omega_{{\bf p}-{\bf q}}}
{\rm e}^{-4p^2/\Lambda^2}
\right]
\nonumber\\
&\times&\left( X^{ij}_{ln}({\bf q})\phi_{+}^{ln}({\bf q})
+Y^{ij}_{ln}({\bf q})\phi_{-}^{ln}({\bf q})
\right)
\nonumber\\
&+&\left[
-\frac{1}{8} N_c \int{d{\bf p}\over(2\pi)^3}\widetilde{V}_{L+C}({\bf
p}-{\bf q})
\frac{(\omega_{\bf p}+\omega_{\bf q})^2}{\omega_{\bf p}\omega_{\bf q}}
\left( X^{ij}_{km}({\bf q})X^{km}_{ln}({\bf p})\phi_{+}^{ln}({\bf p})
+Y^{ij}_{km}({\bf q})Y^{km}_{ln}({\bf p})\phi_{-}^{ln}({\bf p})
\right) \right.
\nonumber\\
&+&\left.
\alpha_s\pi N_c \int{d{\bf p}\over(2\pi)^3}
\frac{2}{\omega_{\bf p}\omega_{\bf q}}
\left( D_{ij}({\bf q})D_{ln}({\bf p})\phi_{+}^{ln}({\bf p})
-\frac{1}{4}X^{ij}_{km}({\bf q})\left[X^{km}_{ln}({\bf
p})\phi_{+}^{ln}({\bf p})
+Y^{km}_{ln}({\bf p})\phi_{-}^{ln}({\bf p})\right]  \right)
\right.
\nonumber\\
&+&\left.
\alpha_s 2\pi N_c\int{d{\bf p}\over(2\pi)^3}
{1\over\omega_{\bf p}\omega_{\bf q}\omega^2_{{\bf p}-{\bf q}} }
\left(1-\frac{(\omega_{\bf p}-\omega_{\bf q})^2}
{(\omega_{\bf p}-\omega_{\bf q})^2+\omega^2_{{\bf p}-{\bf q}}} \right)
\right.
\nonumber\\
&\times&\left. T_{km,k'm'}({\bf q},{\bf p})
\left( X^{ij}_{kk'}({\bf q})X^{mm'}_{ln}({\bf p})\phi_{+}^{ln}({\bf p})
+Y^{ij}_{kk'}({\bf q})Y^{mm'}_{ln}({\bf p})\phi_{-}^{ln}({\bf p}) \right)
\right]
\,,\label{eq}\end{eqnarray}
where the factor $T_{km,k'm'}$ is defined in Eq. (B4)
and
\begin{eqnarray}
&& \phi_{\pm}^{ln}({\bf p})=\frac{1}{2}(\phi^{ln}({\bf p}) \pm
\phi^{ln}(-{\bf p})
)\nonumber\\  && X^{ij}_{ln}({\bf q})=D_{il}({\bf
q})D_{jn}({\bf q})+D_{in}({\bf q})D_{jl}({\bf q})\nonumber\\ &&
Y^{ij}_{ln}({\bf
q})=D_{il}({\bf q})D_{jn}({\bf q})-D_{in}({\bf q})D_{jl}({\bf q})
\,,\label{eq}\end{eqnarray}
involving the polarization sum $D_{ij}({\bf q})$, Eq. (3.13), and glueball
momentum wavefunction, $\phi^{ln}({\bf p})$, introduced in Eq. (4.6).

The bound state equation can be further reduced
by forming the appropriate tensor contractions for $\phi^{ij}({\bf p})$
corresponding to a specific glueball state having quantum numbers
$J$, total angular momentum, $P$, parity,  and $C$,
charge conjugation.  For the applications in this paper
we considered the scalar
($J^{PC} = 0^{++}$) and pseudoscalar ($J^{PC} = 0^{-+}$) channels having tensor
contractions
$\phi^{ij}({\bf q})=\delta^{ij}\phi(q)$ and
$\phi^{ij}({\bf q})=\epsilon^{ijk}{\hat q}^{k}\phi(q)$, respectively.
The resulting equation describing both $J = 0$ states is
\begin{eqnarray}
& & M \phi(q)=
 \left[\left(\frac{q^2+m_{CT}^2(\Lambda)}{\omega_{\bf q}}+\omega_{\bf q}
\right)\right.
\nonumber\\
&+& \left.\frac{1}{4} N_c
\int{d{\bf p}\over(2\pi)^3}
\widetilde{V}_{L+C}({\bf p}-{\bf q})\left(1+({\hat p}{\hat q})^2
\right)
\frac{\omega_{\bf p}^2+\omega_{\bf q}^2}{\omega_{\bf p}\omega_{\bf q}}
{\rm e}^{-p^2/\Lambda^2}  \right.
\nonumber\\
&+&\left. \alpha_s\pi N_c
\int{d{\bf p}\over(2\pi)^3}{1\over\omega_{\bf p}\omega_{\bf q}}
\left(3-({\hat k}{\hat q})^2 \right)
{\rm e}^{-p^2/\Lambda^2}     \right.
\nonumber\\
&-&\left. \alpha_s\pi N_c
\int{d{\bf p}\over(2\pi)^3}{1\over\omega_{\bf p}\omega_{\bf q}\omega_{{\bf
p}-{\bf q}} }
\frac{G({\bf p},{\bf q})}{\omega_{\bf p}+\omega_{{\bf p}-{\bf q}}}
{\rm e}^{-4p^2/\Lambda^2}
\right]\thinspace \phi(q)
\nonumber\\
&+&\left[
-\frac{1}{8} N_c \int{d{\bf p}\over(2\pi)^3}\widetilde{V}_{L+C}({\bf
p}-{\bf q})
\frac{(\omega_{\bf p}+\omega_{\bf q})^2}{\omega_{\bf p}\omega_{\bf q}}
F^{JPC}({\bf p},{\bf q}) \phi(p) \right.
\nonumber\\
&+&\left.
\alpha_s\pi N_c \int{d{\bf p}\over(2\pi)^3}
\frac{1}{2\omega_{\bf p}\omega_{\bf q}}
(3-({\hat p}{\hat q})^2)E^{JPC}({\bf p},{\bf q})\phi(p) \right.
\nonumber\\
&+&\left.
\alpha_s 2\pi N_c\int{d{\bf p}\over(2\pi)^3}
{1\over\omega_{\bf p}\omega_{\bf q} } \frac{D^{JPC}({\bf p},{\bf
q})}{\omega^2_{{\bf p}-{\bf q}} }
\left(1-\frac{(\omega_{\bf p}-\omega_{\bf q})^2}
{(\omega_{\bf p}-\omega_{\bf q})^2+\omega^2_{{\bf p}-{\bf q}}} \right)\phi(p)
\right]
\,,\label{eq:a.1}\end{eqnarray}
with spectroscopic terms $D^{JPC}$, $E^{JPC}$ and $F^{JPC}$ for each
channel given by
\begin{eqnarray}
F^{0++}({\bf p},{\bf q}) &=& 1+({\hat p}{\hat q})^2
\nonumber\\
F^{0-+}({\bf p},{\bf q}) &=& 2({\hat p}{\hat q})
\nonumber\\
E^{0++}({\bf p},{\bf q}) &=& 1
\nonumber\\
E^{0-+}({\bf p},{\bf q}) &=& 0
\nonumber\\
D^{0++}({\bf p},{\bf q}) &=& G({\bf p},{\bf q})
= 2(1-({\hat p}{\hat q})^2)\left( p^2+q^2
+ \frac{p^2q^2}{2({\bf p}-{\bf q})^2}(1+({\hat p}{\hat q})^2)
\right)
\nonumber\\
D^{0-+}({\bf p},{\bf q}) &=& 2p^2q^2 (1-({\hat p}{\hat q})^2)\left(
\frac{2}{pq}+\frac{1}{({\bf p}-{\bf q})^2}({\hat p}{\hat q})
\right)
\,.\label{eq:4.2}\end{eqnarray}
An approximate form for Eq. (C3)
is used (see Eq.
(4.9)) in the main text for numerical calculations.

\begin{table}[ht]
\begin{tabular}{|c|c|c|c|c|} \hline
  $J^{PC}$ & $0^{++}$ & $0^{*++}$ & $0^{-+}$ & $0^{*-+}$ \\ \hline
 calculated (MeV) & $1760$ & $2697$
& $2142$ & $2895$ \\ \hline
lattice data \cite{lattice} (MeV) & $1730(80)$ & $2670(130)$
& $2590(130)$ & $3640(180)$ \\ \hline
\end{tabular}
\vspace{.5 cm}
\caption{Glueball spectrum for the first two scalar and pseudoscalar states
($\alpha_s= \frac{g^2}{4\pi} =0.4, \, \sigma=0.18 \, GeV^2, \, \Lambda=4 \,
GeV$).}
\label{tab.1}
\end{table}

\newpage

\newpage

\end{document}